\documentclass[12pt]{article}

\input{paper_preamble.sty}

\title{\vspace{-2.2cm}\Large\bf The Non-Relativistic Geometric Trinity of Gravity}

\author[1]{{William J.} {Wolf}\footnote{\href{mailto:william.wolf@philosophy.ox.ac.uk}{william.wolf@philosophy.ox.ac.uk}}}
\author[1]{{James} {Read}\footnote{\href{mailto:james.read@philosophy.ox.ac.uk}{james.read@philosophy.ox.ac.uk}}}
\author[2]{{Quentin} {Vigneron}\footnote{\href{mailto:quentin.vigneron@umk.pl}{quentin.vigneron@umk.pl}}}

\vspace{-.3cm}
\affil[1]{\small\it{Faculty of Philosophy, University of Oxford\\
Oxford OX2 6GG, United Kingdom}}

\affil[2]{\small\it{Institute of Astronomy, Faculty of Physics, Astronomy and Informatics}, {Nicolaus Copernicus University}, {{Grudzi{\k{a}}dzka 5}, {Toru\'n}, {87-100} {Poland}}}

\date{\vspace{-.3cm}\today\vspace{-.cm}}

\begin{document}

\maketitle

\vspace{-.8cm}
\begin{abstract}
\noindent The geometric trinity of gravity comprises three distinct formulations of general relativity: (i) the standard formulation describing gravity in terms of spacetime curvature, (ii) the teleparallel equivalent of general relativity describing gravity in terms of spacetime torsion, and (iii) the symmetric teleparallel equivalent of general relativity (STEGR) describing gravity in terms of spacetime non-metricity. In this article, we complete a geometric trinity of non-relativistic gravity, by (a) taking the non-relativistic limit of STEGR to determine its non-relativistic analogue, and (b) demonstrating that this non-metric theory is equivalent to the Newton--Cartan theory and its teleparallel equivalent, i.e., the curvature and the torsion based non-relativistic theories that are both geometrised versions of classical Newtonian gravity.

\end{abstract}

\vspace{.05cm}
\hrule

\vspace{-.55cm}
\small
\renewcommand{\contentsname}{\ \vspace{-.5cm}}
\tableofcontents
\normalsize

\hfill 
{\color{white} blank}

\section{Introduction}

It has become increasingly well-known that general relativity (GR) constitutes but one vertex in a `geometric trinity' of gravitational theories \cite{2019_Jimenez_et_al, Capozziello:2022zzh}. The other two vertices of this trinity are the `teleparallel equivalent general relativity' (TEGR), in which the curvature degrees of freedom of GR are traded for spacetime torsion, and the `symmetric teleparallel equivalent general relativity' (STEGR), in which the curvature degrees of freedom of GR (and torsion degrees of freedom of TEGR) are traded for spacetime non-metricity. The actions of all three theories are equivalent up to a total divergence term---in this sense, all three theories are dynamically equivalent.\saut

In a parallel vein, it has been known since Trautman in the 1960s \cite{1965_Trautman} that standard Newtonian gravity can be formulated similarly to GR in the sense that (non-relativistic) gravitational effects become a manifestation of spacetime curvature: this theory is known as Newton--Cartan (NC) theory, and was first developed in the 1920s by Cartan and Friedrichs: see \cite{1923_Cartan, 1924_Cartan, 1928_Friedrichs} for the original sources, and \cite{2023_Hartong_et_al} for a recent review of non-relativistic gravity. (To be clear: in this article, by `non-relativistic', we always mean `Newtonian', rather than `ultra-relativistic'; we will leave the construction of an ultra-relativistic geometric trinity to future work.) In \cite{2018_Read_et_al}, it was
shown that there is a precise sense in which classical Newtonian gravitation can be understood as the teleparallelised version of NC, which we will call TENC, in which the gravitational potential can be understood as a manifestation of the `mass torsion' which arises once one gauges the Bargmann algebra (e.g.,~\cite{2015_Geracie_et_al}). Adding to this, it was shown recently in \cite{2023_Schwartz} that TENC can be secured as the non-relativistic limit of TEGR using a $1/c$ expansion of the TEGR field equation (in \cite{2018_Read_et_al} the same result was shown using null reduction), just as NC is by now well-known to be the non-relativistic limit of GR (on which see \cite{2023_Hartong_et_al} and references therein).\saut


These results invite the following question: \emph{can one complete a non-relativistic geometric trinity}, by constructing a non-metric theory equivalent to both NC (understood as a theory with spacetime curvature) and TENC (understood as a torsionful theory)? In this article, we answer this question in the affirmative---indeed, we triangulate a non-relativistic version of Newtonian gravity (which we dub `symmetric teleparallel equivalent Newton--Cartan' in analogy with its relativistic parent, shortened to STENC) in two ways: (a) by taking the non-relativistic limit of STEGR (using the same $1/c$ expansion  developed in~\cite{1976_Kunzle}), and (b) by proving that it includes classical Newtonian gravity, thereby obtaining the analogues of the Trautman recovery theorems (see \cite[Ch.~4]{2012_Malament}) for STENC.\saut

The structure of this article is as follows. In Sec.~\ref{sec:trinity}, we review the essential details of the geometric trinity of gravity; in Sec.~\ref{sec:limit}, we construct STENC by taking the non-relativistic limit of STEGR and show that it is equivalent to the NC formulation; in Sec.~\ref{sec_discussion}, we discuss general properties of the non-relativistic trinity and of STENC specifically. We close in Sec.~\ref{sec:close} with some discussions of the upshots of this work.

\section{Background: The Geometric Trinity}\label{sec:trinity}

The bulk of this section constitutes a review of the relativistic geometric trinity of gravity (Sec.~\ref{sec:review}). In addition, we review briefly the state-of-play regarding geometric reformulations of non-relativistic gravity (Sec.~\ref{sec:nonrel}).

\subsection{Notation}\label{subsec:notation}

Since we will deal with four different connections, introduced at the relativistic and the non-relativistic levels, we here define the following notation:
\begin{enumerate}[label=(\roman*)]
    \item The Lorentzian Levi-Civita connection $\Gamma^\mu{}_{\alpha\beta}$ relative to the Lorentzian metric $g_{\mu\nu}$, with covariant derivative $\nabla$, and Riemann tensor $R^\mu{}_{\alpha\beta\nu}$.

    \item The general affine connection introduced at the relativistic level $\gene\Gamma^\mu{}_{\alpha\beta}$, with covariant derivative $\gene\nabla$, torsion $\geneT{T}^\mu{}_{\alpha\beta}$, and Riemann tensor $\gene R^\mu{}_{\alpha\beta\nu}$.

    \item The symmetric Galilean connection $\hat \Gamma^\mu{}_{\alpha\beta}$ (i.e., non-relativistic connection) relative to $(\tau_\mu, h^{\mu\nu})$, with covariant derivative $\hat\nabla$, and Riemann tensor $\hat R^\mu{}_{\alpha\beta\nu}$. For simplicity, we will refer to $\hat \Gamma^\mu{}_{\alpha\beta}$ as a `Galilean connection', dropping the `symmetric' in the name.

    \item The general affine connection introduced at the non-relativistic level $\geneNR\Gamma^\mu{}_{\alpha\beta}$, with covariant derivative $\geneNR\nabla$, torsion $\geneNRT T^\mu{}_{\alpha\beta}$, and Riemann tensor $\geneNR R^\mu{}_{\alpha\beta\nu}$.

\end{enumerate}
When considering the non-relativistic limit, both the Lorentzian connection $\Gamma^\mu{}_{\alpha\beta}$ (and related variables) and the general affine connection $\gene\Gamma^\mu{}_{\alpha\beta}$ (and related variables) are written as Taylor series of the speed of light. The full series will be denoted by an upper ``$\lambda$'', where $\lambda = 1/c^2$ (e.g., $\accentset{\lambda}{\Gamma}^\mu{}_{\alpha\beta}$), and the $n$-th order will be denoted by an upper ``$(n)$'' (e.g., $\tayl{n}{\Gamma}^\mu{}_{\alpha\beta}$).\saut\saut

Finally, as will be detailed in Sec.~\ref{sec:nonrel}, a non-relativistic (Galilean) structure does not possess a metric allowing for raising and lowering indices. Therefore, when introducing tensors at the non-relativistic level (denominated with a hat), the position of the indices will be fixed.\\

\begin{remark}
    Here, we draw attention to the use of the reference connection $\bar\nabla$, where this denotes a general affine connection with possible torsion and/or non-metricity.  As this article is already relatively heavy with notation between relativistic and non-relativistic connections, we will not denote separate connections for a torsionful or non-metric connection; rather, we will flag instances where the torsion and/or non-metricity have been turned off as we do in e.g.~Eqs.~\eqref{eq_TEGR} and \eqref{eq_STEGR} where we explicitly specify that in TEGR and STEGR the respective non-metricity and torsion have been switched off for those specific theories. This choice of notation carries over to the non-relativistic level, where $\bar\Gamma$ is replaced by $\tilde\Gamma$ for TENC and STENC.
\end{remark}

\subsection{Relativistic gravity}\label{sec:review}

Spacetime theories are typically formulated in terms of a metric tensor $g_{\mu\nu}$ and an affine connection $\gene{\Gamma}^{\alpha}{}_{\mu\nu}$. General relativity (GR) is of course the paradigmatic theory of gravity and makes use of the Levi-Civita connection $\Gamma^\alpha{}_{\mu\nu}$, setting $\gene\Gamma^\mu{}_{\alpha\beta} = \Gamma^\mu{}_{\alpha\beta}$, with components
\begin{equation}
    \Gamma^\alpha{}_{\mu \nu}:=\frac{1}{2} g^{\alpha \lambda}\left(g_{\lambda \nu, \mu}+g_{\mu \lambda, \nu}-g_{\mu \nu, \lambda}\right),
\end{equation}
which is the unique connection that is compatible with the metric and torsion-free. The metric-compatibility condition is given by $\nabla_{\alpha}g_{\mu\nu} = 0$ and the torsion-free condition is given by $\Gamma\indices{^{\alpha}_{[\mu \nu]}}=0$ \cite[Ch.~3]{1984_Wald}. Famously, GR describes gravity as a manifestation of spacetime curvature, as encoded in the Riemann tensor, which has components
\begin{equation}
    R\indices{^\alpha_{\beta \mu \nu}}(\Gamma):=
        \partial_\mu \Gamma\indices{^\alpha_{\nu \beta}}-\partial_\nu \Gamma\indices{^\alpha_{\mu \beta}} + \Gamma\indices{^\alpha_{\mu \lambda}} \Gamma\indices{^\lambda_{\nu \beta}}-\Gamma\indices{^\alpha_{\nu \lambda}} \Gamma\indices{^\lambda_{\mu \beta}}.
\end{equation}
Spacetime curvature measures the rotation of a vector when it is parallel transported along a closed curve.\saut

One can alter or otherwise relax the above assumptions in order to construct spacetime theories that manifest torsion and/or non-metricity. Torsion is given by the antisymmetric part of the connection 
\begin{equation}
\bar T\indices{^\alpha_{\mu \nu}}\left(\gene{\Gamma}^\alpha{}_{\mu\nu}\right) := 2 \gene{\Gamma}\indices{^\alpha_{[\mu \nu]}},
\end{equation}
and can be thought of as a measure of the non-closure of the infinitesimal parallelogram formed by two vectors being parallel transported along each other. Non-metricity is given by the non-vanishing of the covariant derivative of the metric tensor
\begin{equation}
Q_{\alpha \mu \nu}\left(\gene{\Gamma}^\alpha{}_{\mu\nu}\right) := \gene{\nabla}_\alpha g_{\mu \nu},
\end{equation}
and can be thought of as a measure of how the length of a vector changes when parallel transported.\saut

We can thus categorize spacetimes as 
\begin{enumerate}[label=(\roman*)]
    \item metric (i.e.,\ $Q_{\alpha \mu \nu}\left(\gene{\Gamma}^\alpha{}_{\mu\nu}\right) = 0$),
    \item torsionless (i.e.,\ $\bar T\indices{^\alpha_{\mu \nu}}\left(\gene{\Gamma}^\alpha{}_{\mu\nu}\right) =0$),
    \item flat (i.e.,\ $R\indices{^\alpha_{\beta \mu \nu}}\left(\gene{\Gamma}^\alpha{}_{\mu\nu}\right) =0$).
\end{enumerate}
Curvature, torsion, and non-metricity are all possible geometric properties of an affine connection in relation with a Lorentzian metric. A completely general affine connection $\gene{\Gamma}\indices{^\alpha_{\mu \nu}}$ can be decomposed in the following way \cite{2004_Ortin}:
\begin{equation}\label{decomp}
    \gene{\Gamma}\indices{^\alpha_{\mu \nu}}=\Gamma\indices{^\alpha_{\mu \nu}}+K\indices{^\alpha_{\mu \nu}}+L\indices{^\alpha_{\mu \nu}},
\end{equation}
where
\begin{equation}\label{eq_def_K_GR}
K\indices{^\alpha_{\mu \nu}} := \frac{1}{2} \geneT{T}^\alpha{}_{\mu\nu} + g^{\alpha \sigma} \geneT{T}^\gamma{}_{\sigma (\mu} \, g_{\nu)\gamma}
\end{equation}
is referred to as the `contortion tensor', and
\begin{equation}\label{eq_def_L_GR}
L\indices{^\alpha_{\mu \nu}}:=\frac{1}{2} g^{\alpha \sigma}Q_{\sigma \mu \nu} - g^{\alpha \sigma} {Q}_{(\mu\nu) \sigma}
\end{equation}
is referred to as the `distorsion tensor'. Consequently, the difference tensor $\Gamma^\mu{}_{\alpha\beta} - \gene{\Gamma}^\mu{}_{\alpha\beta}$ between the Levi-Civita connection and a general affine connection has the form
\begin{align}
    \Gamma^\mu{}_{\alpha\beta} - \gene{\Gamma}^\mu{}_{\alpha\beta} = 
        g^{\mu\sigma}\left(\gene{\nabla}_{(\alpha} g_{\beta)\sigma} - \frac{1}{2} \gene{\nabla}_{\sigma} g_{\alpha\beta} \right) - \frac{1}{2} \geneT{T}^\mu{}_{\alpha\beta} - g^{\mu \sigma} \geneT{T}^\nu{}_{\sigma (\alpha} \, g_{\beta)\nu}. \label{eq_gene_diff_Gamma}
\end{align}
This formula, equivalent to (8), will be useful when taking the non-relativistic limit in Sec.~\ref{sec:limit}.

We can use Eq.~\eqref{decomp} to facilitate translations between different spacetime theories with different connections (and associated different geometrical properties).
As we have seen, GR is a spacetime theory that is metric and torsionless, but non-flat as the Levi-Civita connection in general possesses curvature. In this article, we will also be concerned with two other spacetime theories: the `teleparallel equivalent general relativity' (TEGR) and the `symmetric teleparallel equivalent general relativity' (STEGR). The connection of TEGR is metric and flat but in general possess torsion; the connection of STEGR is torsionless and flat but in general possess non-metricity. Both TEGR and STEGR are dynamically equivalent to GR, in the sense that the actions of all three theories are equivalent up to total divergence terms [see Eqs~\eqref{eq_Lag_R}--\eqref{eq_Lag_T}]; thereby, these theories are capable of modelling the same empirical phenomena, and constitute a `geometric trinity' of gravity---see \cite{2019_Jimenez_et_al, 2019_Heisenberg, 2023_Capozziello_et_al, 2023_Wolf_et_al_b, Nester:1998mp, 2012_Aldrovandi_et_al} for further discussions.\saut

For example, we can find GR's torsionful and non-metric equivalents by taking the expressions for the Riemann curvature and Ricci scalar in GR in terms of the Levi-Civita connection, and re-expressing these in terms of the torsionful connection of TEGR or the non-metric connection of STEGR. Consider that we can express a generic Riemann curvature tensor $\gene{R}\indices{^\alpha_{\beta \mu \nu}}$ as \cite{2018_Jarv_et_al}:
\begin{align}\label{eq:RR}
    \gene{R}\indices{^\alpha_{\beta \mu \nu}} =  R\indices{^\alpha_{\beta \mu \nu}} +\nabla_\mu M\indices{^\alpha_{\nu\beta}}-\nabla_\nu M\indices{^\alpha_{\mu\beta}} 
    +M\indices{^\gamma_{\nu \beta}} M\indices{^\alpha_{\mu \gamma}}-M\indices{^\gamma_{\mu \beta}} M\indices{^\alpha_{\nu \gamma}} + \geneT T\indices{^\alpha_{\beta \mu}} M\indices{^\beta_{\alpha \nu}},
\end{align}
where $R\indices{^\alpha_{\beta \mu \nu}}$ is the standard Riemann tensor from the Levi-Civita connection $\nabla_\mu$ and $M\indices{^\alpha_{\mu \nu}} := K\indices{^\alpha_{\mu \nu}}+L\indices{^\alpha_{\mu \nu}}$. This formula is the heart of the trinity of GR. 
One can choose to work with TEGR and the contortion tensor (i.e., $K\indices{^\alpha_{\mu \nu}}\neq 0$ and  $L\indices{^\alpha_{\mu \nu}} =0$) or with STEGR and the distorsion tensor (i.e., $K\indices{^\alpha_{\mu \nu}} = 0$ and  {${L\indices{^\alpha_{\mu \nu}} \neq 0}$}). Upon index contraction, one constructs the curvature scalar and finds: 
\begin{align}\label{eq_diff_Div_T}
    - R &= \frac{1}{4} \geneT T_{\alpha\mu\nu} \geneT T^{\alpha\mu\nu} + \frac{1}{2} \geneT T_{\alpha\mu\nu}\geneT T^{\mu\alpha\nu} - \geneT T^\alpha{}_{\alpha\mu}\geneT T^\beta{}_{\beta\nu}g^{\mu\nu} + 2 \nabla_\alpha \geneT T\indices{^\lambda^\alpha_\lambda} \quad {\rm with} \quad Q_{\mu\alpha\beta} = 0,
\end{align}
and
\begin{align}\label{eq_diff_Div_Q}
    - R &= g^{\mu\nu} \left(L^\alpha{}_{\mu\beta}L^\beta{}_{\nu\alpha} - L^\alpha{}_{\alpha\beta}L^\beta{}_{\mu\nu}\right) +\nabla_\alpha\left(Q\indices{^\alpha_\lambda^\lambda}-Q\indices{^\lambda_\lambda^ \alpha}\right)  \quad {\rm with} \quad \geneT T^\alpha{}_{\mu\nu} = 0.
\end{align}
Importantly, this shows that the scalar expressions of curvature, torsion, and non-metricity are equivalent up to a boundary term.\footnote{See \cite{2017_Oshita_et_al, 2022_Heisenberg_et_al, 2023_Wolf_et_al_a} for some discussions concerning the role and significance of these boundary terms.} This justifies the above claim that GR, TEGR, and STEGR can be formulated in terms of dynamically equivalent Lagrangian expressions, respectively as
\begin{align}
	\CL_R &\coloneqq g^{\mu\nu} R_{\mu\nu}, \label{eq_Lag_R} \\
	\CL_T &\coloneqq \frac{1}{4} \bar T_{\alpha\mu\nu}\bar T^{\alpha\mu\nu} + \frac{1}{2} \bar T_{\alpha\mu\nu}\bar T^{\mu\alpha\nu} - \bar T^\alpha{}_{\alpha\mu}\bar T^\beta{}_{\beta\nu}g^{\mu\nu} + \left(\lambda_\mu{}^{\nu\alpha\beta} \bar R^\mu{}_{\nu\alpha\beta} 
    + \lambda^{\mu\nu\alpha} Q_{\mu\nu\alpha}\right), \label{eq_Lag_T} \\
	\CL_Q &\coloneqq g^{\mu\nu} \left(L^\alpha{}_{\mu\beta}L^\beta{}_{\nu\alpha} - L^\alpha{}_{\alpha\beta}L^\beta{}_{\mu\nu}\right) + \left(\lambda_\mu{}^{\nu\alpha\beta} \bar R^\mu{}_{\nu\alpha\beta} 
    + \lambda^{\mu\nu\alpha} \bar T_{\mu\nu\alpha}\right), \label{eq_Lag_Q}
\end{align}
where the $\lambda$ tensors are Lagrange multipliers setting the reference curvature and the non-metricity (or torsion) to zero. 
Note that we could remove the need for these multipliers by writing $\bar\nabla$ as a pure gauge connection (see \citep{2019_Jimenez_et_al}), or by choosing a specific gauge (e.g. ${\bar\Gamma^\mu{}_{\alpha\beta} = 0}$). We choose not to do so in this article in order to avoid the need to introduce additional variables such as tetrads which might impair the readability of the paper.\footnote{This is especially relevant since we will focus mainly on the non-relativistic equivalent of STEGR and that the non-relativistic equivalent of TEGR has already been developed in \citep{2023_Schwartz}, in which tetrads were used.}

From the Lagrangians~\eqref{eq_Lag_R}-\eqref{eq_Lag_Q}, the same equation of motion is found, but written as a function of either the Levi-Civita Ricci curvature $R_{\mu\nu}$, the contortion $K^\mu{}_{\alpha\beta}$, or the distorsion $L^\mu{}_{\alpha\beta}$:
\begin{gather}
    {\rm GR:} \ 
    R_{\mu\nu} = \frac{8\pi G}{c^4}\left(T_{\mu\nu} - \frac{T^\alpha{}_\alpha}{2}g_{\mu\nu}\right)\\
	\Updownarrow  \nonumber\\
    {\rm TEGR:} \ 
    \begin{dcases} \label{eq_TEGR}
        &-\gene\nabla_\alpha  K^\alpha{}_{\mu\nu} + \gene\nabla_\mu K^\alpha{}_{\nu\alpha} - K^\alpha{}_{\mu\beta} K^\beta{}_{\alpha\nu} + K^\alpha{}_{\alpha\beta} K^\beta{}_{\mu\nu} \\
        &\qquad\qquad\qquad\qquad\qquad\qquad+ \ \geneT T^\alpha{}_{\beta\mu}K^\beta{}_{\alpha\nu} = \frac{8\pi G}{c^4}\left(T_{\mu\nu} - \frac{T^\alpha{}_\alpha}{2}g_{\mu\nu}\right), \\
        &{\rm with} \quad \gene R^\mu{}_{\alpha\beta\sigma} = 0 \quad {\rm and} \quad Q_{\mu\alpha\beta} = 0,
    \end{dcases}\\
	\Updownarrow  \nonumber \\
    {\rm STEGR:} \ 
    \begin{dcases} \label{eq_STEGR}
        &-\bar\nabla_\alpha  L^\alpha{}_{\mu\nu} + \bar\nabla_\mu  L^\alpha{}_{\nu\alpha} -  L^\alpha{}_{\mu\beta}  L^\beta{}_{\alpha\nu} +  L^\alpha{}_{\alpha\beta} L^\beta{}_{\mu\nu} = \frac{8\pi G}{c^4}\left(T_{\mu\nu} - \frac{T^\alpha{}_\alpha}{2}g_{\mu\nu}\right), \\
        &{\rm with} \quad \gene R^\mu{}_{\alpha\beta\sigma} = 0 \quad {\rm and} \quad \geneT T^\alpha{}_{\mu\nu} = 0,
    \end{dcases}
\end{gather}
where $T_{\mu\nu}$ is the energy momentum tensor. These equivalences are a direct consequence of relation~\eqref{eq:RR}, relating the Riemann tensors of two different connections.\saut

While these particular theories are empirically equivalent to each other, there are a number of reasons why physicists are interested in investigating such alternative geometric representations. One reason has to do with the fact that these theories possess different gauge structure. In particular, TEGR and STEGR can be understood as gauge theories of translations \cite{2017_Jimenez_et_al, 2012_Aldrovandi_et_al}, which allows one to formulate the theories in a language more closely resembling other fundamental interactions and potentially suggests different routes towards quantisation. Another reason can be found in resolving cosmological puzzles. Despite the incredible successes of the current $\Lambda$CDM model, there are a number of unresolved issues that are the subject of heated debate, including our modeling of both early and late time expansion of the universe \cite{2016_Joyce_et_al, 2023_Wolf_et_al_c, 2017_Brandenberger_et_al, 2019_Chowdhury_et_al, 2022_Wolf_et_al}. While the theories within the trinity are indeed equivalent, their geometric structures based on curvature, torsion, and non-metricity suggest different routes to modifying gravity. Indeed, the equivalence is broken when we move to modifications that consist in higher order scalar invariants of the relevant geometric quantities. That is, e.g.,~$f(R)$, $f(T)$, and $f(Q)$ theories are not equivalent to each other, and this has motivated exploring this theory space as possible novel realisations of dark energy, inflation, the astrophysics of black holes, and bouncing cosmologies \cite{2023_Bahamonde_et_al, 2011_Cai_et_al, 2020_Bajardi_et_al, 2020_Jimenez_et_al, 2010_Linder}.\\

Note that in formulating this geometric trinity of gravity, we are following the lead of \cite{2019_Jimenez_et_al, Nester:1998mp, 2017_Jimenez_et_al, 2020_Jimenez_et_al, Capozziello:2022zzh} and use the metric formalism rather than the tetrad formalism. 
We emphasize that to work in the metric formulation rather than the tetrad formulation is perfectly legitimate as both formalisms are physically equivalent; however, it is worth noting that insofar as one might seek to study the status of the nodes of the trinity as gauge theories of gravity, it might be more congenial to work with tetrads and spin connections---see e.g.\ 
\cite{2012_Aldrovandi_et_al, 1995_Hehl_et_al} in the relativistic case. We expect this to be true of the non-relativistic geometric trinity also, and indeed see \cite{Schwartz_2024aee} for some initial explorations in this direction using the formalism of tetrads and spin connections. However, since our primary concern in this article is the \emph{construction} of the non-relativistic geometric trinity, it suffices for us to use the formalism of metrics and affine connections. Furthermore, much of this article concerns taking the non-relativistic limit of STEGR, which as a theory whose gravitational degrees of freedom are encoded in its non-metricity, is primarily presented using the metric formalism (e.g.~see the discussion in \cite[Sec.~B.4]{Capozziello:2022zzh}). Working in this formalism allows us to more clearly align with the prevailing literature on this particular node of the trinity as well as render transparent the geometric degrees of freedom present in this theory.

\subsection{Non-relativistic gravity}\label{sec:nonrel}

So much by way of background on the relativistic geometric trinity of gravity; what is the current state-of-the-art with respect to non-relativistic physics? It has been known since the 1960s that standard, flat-space classical Newtonian gravity can be formulated as a curved spacetime theory known as `Newton--Cartan theory' (NC), in a way closely resembling GR. The geometric structure involved in the NC formulation is known as a `Galilean structure', defined by degenerate spatial and temporal metrics $h^{\mu\nu}$ and $\tau_{\mu}$ (assumed in this paper to be exact) that are orthogonal, i.e. $\tau_{\mu} h^{\mu\nu} \coloneqq 0$, and equipped with a Galilean connection $\hat{\Gamma}^{\alpha}{}_{\mu\nu}$ that is separately compatible with both spatial and temporal metrics such that $\hat{\nabla}_{\alpha} h^{\mu\nu} \coloneqq 0$ and $\hat{\nabla}_{\alpha} \tau_{\mu} \coloneqq 0$. 
Similarly to GR, the dynamical degrees of freedom of NC theory are captured by its curvature tensor $\hat{R}\tensor{\vphantom{R}}{^\alpha_\beta_\mu_\nu}$, with the field equations
\begin{align}
    \hat{R}\tensor{\vphantom{R}}{_\mu_\nu}              &= 4\pi\rho \tau_{\mu} \tau_{\nu}, \qquad \\
    \hat{R}\tensor{\vphantom{R}}{^\alpha_\nu^\mu_\beta} &= \hat{R}\tensor{\vphantom{R}}{^\mu_\beta^\alpha_\nu}, \qquad \\
        \hat\nabla_\mu T^{\mu\nu}                       &= 0,
\end{align}
where $\rho$ is the mass density. The first equation is known as the Newton--Cartan equation, and encodes, in particular,  the Newton-Poisson equation (see e.g.~\cite[Ch.~4]{2012_Malament}). The second equation is a condition (due to Trautman) that the curvature tensor must satisfy  such that inertial frames, a defining feature of Newtonian theory, exist. The third equation is the conservation of the energy-momentum tensor. Contrary to the GR case, the first equation does not implies the third, which, therefore, is an additional independent equation.\saut

The equivalence of this spacetime theory with classical Newtonian gravity is most easily obtained by projecting along $h^{\mu\nu}$ and $\tau_\mu$ the above system of equations \citep{2021_Vigneron}, i.e., performing a 3+1-projection. Recovering classical Newtonian theory is also codified in the Trautman geometrisation and recovery theorems \cite[Ch.~4]{2012_Malament}, where one can define another connection~$\nabla'$ such that $\hat{\nabla}_\beta v^\alpha=\nabla'_\beta v^\alpha - \tau_\beta \tau_{\nu} v^\nu h^{\alpha\mu} \nabla'_\mu \Phi$ where $\Phi$ is the familiar Newtonian gravitational potential. Upon assuming an isolated system, i.e., no cosmic expansion, one can then show that this connection is also compatible with the metrics and leads to a flat spacetime where one recovers the familiar Poisson equation $\nabla'^\alpha \nabla'_\alpha  \Phi = 4\pi\rho$.\saut

The Newton--Cartan formulation, and non-relativistic gravity more generally, is still a very active field of research. It has found important theoretical applications in non-relativistic holography, non-relativistic string theory, and quantum gravity \cite{2016_Hartong_et_al, Bergshoeff_2019pij, 2011_Andringa_et_al, 2015_Hartong_et_al, Bergshoeff_2014uea} because non-relativistic gravity often provides a more controlled environment where one can gain greater computational traction, and this can facilitate exploration and understanding of the more complicated relativistic counterparts. Furthermore at a more emergent level, non-relativistic gravity has been incredibly useful in modelling condensed matter systems such as the fractional quantum Hall effect \cite{2013_Son, 2015_Geracie_et_al, 2021_Wolf_et_al, 2012_Bekaert_et_al} and non-relativistic fluids and hydrodynamics \cite{2014_Jensen,  2015_Geracie_etal2}.
A critical realisation is that, despite the utility of General Relativity and relativistic concepts in so many areas of physics, there are many emergent non-relativistic systems which do not have the same relativistic symmetries. Consequently, non-relativistic gravity becomes a crucial tool with which to study them because it has the appropriate symmetries (see e.g.\ \cite{2013_Son, 2021_Wolf_et_al} for some discussions).\saut

Furthermore, introducing torsion in the Newton--Cartan framework has been shown to be crucial for spurring many of these developments (see e.g.\ \cite{march_read_teh_wolf_2024} for an overview of some of the recent literature).
In particular, as explained in \cite{2019_Hansen_et_al_b, 2015_Hartong_et_al}, the presence of mass in non-relativistic theories necessarily implies the non-vanishing of temporal torsion. Similarly (and relatedly), \cite[Sect.~II.C]{2014_Geracie_etal} explains that conservation of energy requires the introduction of non-zero temporal torsion in non-relativistic gravity and that temporal torsion is needed to properly define an energy current for such non-relativistic systems. Additionally, the existence of torsion is important due to symmetry considerations in non-relativistic holography \cite{Bergshoeff_2014uea}.\saut

What constitutes {very} recent knowledge in torsional Newton--Cartan theory is that it is possible to obtain a teleparallel equivalent Newton--Cartan theory (TENC) \citep{2018_Read_et_al, 2023_Schwartz}, similarly as TEGR for relativistic gravity. In this TENC formulation, the gravitational field present in the classical Newtonian formulation can be understood as the torsion of the mass gauge field $m_\mu$ obtained by gauging the Bargmann algebra \cite{2018_Read_et_al}. In other words, the torsion of the flat-torsionful-metric connection $\geneNR\nabla$ present in TENC is sourced by the mass gauge field. Moreover, TENC can be obtained by taking a $1/c^2$ expansion of TEGR \cite{2023_Schwartz}.\saut

\begin{figure}[t]
\centering
\begin{tikzcd}
\text{GR} \arrow[rr, Leftrightarrow]
 \arrow[dr, Leftrightarrow] \arrow[dd,"c \rightarrow \infty" left]
&& \text{TEGR} \arrow[dl, Leftrightarrow] \arrow[dd,  "c \rightarrow \infty" right,] \\
& \text{STEGR}  & \\
\text{NC} \arrow[rr, Leftrightarrow] \arrow[dr, Leftrightarrow, dashed]
&& \text{TENC} \arrow[dl, Leftrightarrow, dashed] \\
& \text{STENC} \arrow[uu, leftarrow, dashed, "c \rightarrow \infty"  {xshift=26pt, yshift=11pt}, crossing over]&
\end{tikzcd}
\caption{The geometric trinity and its (conjectured) non-relativistic limit.}
\label{fig:1}
\end{figure}
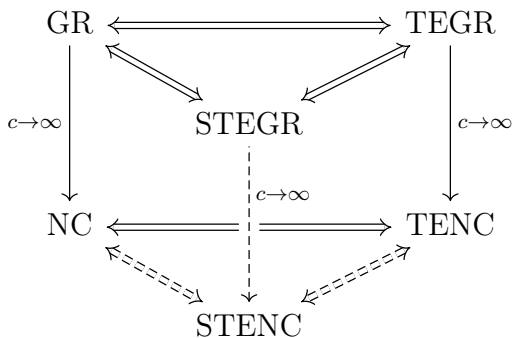

Therefore, in view of the relativistic trinity presented in Sec.~\ref{sec:review} and the aforementioned developments, this invites the following questions: (a) can one construct a non-metric non-relativistic theory of gravity by taking a $1/c^2$ expansion of STEGR, and (b) is that theory equivalent to the NC theory, thereby retrieving the classical Newtonian theory?
In the remainder of this article, we answer in the affirmative both (a) and (b), obtaining, as a result, a symmetric teleparallel equivalent Newton--Cartan theory (STENC). Thereby, we fill in the dotted lines in the Figure \ref{fig:1}, and so complete for the fist time a non-relativistic geometric trinity of gravity, which we summarise in Sec.~\ref{sec_NRTrinity}.

\section{The Non-Relativistic Limit of the Trinity}\label{sec:limit}

The goal of this section is to perform the non-relativistic limit of the STEGR field equations. For this, we will first define what is, in general, a contortion $\hat K^\mu{}_{\alpha\beta}$ and a distorsion~$\hat L^\mu{}_{\alpha\beta}$ relative to a Galilean connection $\hat\nabla$ (Sec.~\ref{sec_dis_cont_NR}). We will then define the non-relativistic limit using Lorentzian and Galilean structures (Sec.~\ref{sec_NR_limit}), and apply this limit to a relativistic contortion $K^\mu{}_{\alpha\beta}$, distorsion $L^\mu{}_{\alpha\beta}$ and difference tensor $\Gamma^\mu{}_{\alpha\beta} - \gene\Gamma^\mu{}_{\alpha\beta}$ (Sec.~\ref{sec_cont_dis_NR_limit}). Finally, in Sec.~\ref{sec_GR_NR_limit}, we perform the limit of the relativistic trinity, hence rederiving the TENC formulation, but most importantly, obtaining the STENC formulation.

\subsection{Distorsion and contortion tensors with respect to a Galilean connection} \label{sec_dis_cont_NR}

In this section, we define the notion of distorsion $\hat L^\mu{}_{\alpha\beta}$ and contortion $\hat K^\mu{}_{\alpha\beta}$ relative to a Galilean connection $\hat \Gamma^\mu{}_{\alpha\beta}$ and a general affine connection $\geneNR\Gamma^\mu{}_{\alpha\beta}$.\saut

Contrary to the Levi-Civita connection of a Lorentzian metric, a Galilean connection defined with respect to a spatial metric $h^{\mu\nu}$ and a temporal metric $\tau_\mu$ is not unique. Two freedoms exist: (i) in the choice of a timelike vector $B^\mu$ defined such that $B^\mu \tau_\mu = 1$; (ii)~in the choice of a 2-form $\kappa_{\mu\nu}$ called the Coriolis field. The connection coefficients $\hat\Gamma^\mu{}_{\alpha\beta}$ have the form
\begin{align} \label{eq_Galilean_connection}
    \hat\Gamma^\mu{}_{\alpha\beta} = h^{\mu\sigma}\left(\partial_{(\alpha} b_{\beta)\sigma} - \frac{1}{2} \partial_{\sigma} b_{\alpha\beta} \right) + B^\mu \partial_{(\alpha} \tau_{\beta)} + 2\tau_{(\alpha} \kappa_{\beta)\nu}  h^{\mu\nu},
\end{align}
where $b_{\mu\nu}$ is the spatial projector orthogonal to $B^\mu$ defined such that $b_{\mu\nu} B^\mu \coloneqq 0$ and $h^{\alpha\mu}b_{\beta\mu} \coloneqq \delta^\alpha_\beta - \tau_\beta B^\alpha$. For a given Galilean connection, $\hat\Gamma^\mu{}_{\alpha\beta}$ the tensors $B^\mu$ and $\kappa_{\mu\nu}$ are not unique and are related by $\kappa_{\mu\nu} = b_{\sigma[\nu} \hat\nabla_{\mu]} B^\sigma$ (see Appendix~\ref{app_Coriolis} for more precision).\saut

We now consider an additional general affine connection $\geneNR{\Gamma}^\alpha{}_{\mu\nu}$ whose torsion is denoted {$\geneNRT T^\alpha{}_{\mu\nu} \coloneqq 2 \geneNR{\Gamma}^\alpha{}_{[\mu\nu]}$}. Using the general formula~\eqref{eq_Galilean_connection} for a symmetric Galilean connection, the difference tensor $\hat\Gamma^\mu{}_{\alpha\beta} - \geneNR{\Gamma}^\mu{}_{\alpha\beta}$ takes the form
\begin{align}
    \hat\Gamma^\mu{}_{\alpha\beta} - {\geneNR{\Gamma}}^\mu{}_{\alpha\beta} &= 
        h^{\mu\sigma}\left({\geneNR{\nabla}}_{(\alpha} b_{\beta)\sigma} - \frac{1}{2} {\geneNR{\nabla}}_{\sigma} b_{\alpha\beta} \right) + B^\mu {\geneNR{\nabla}}_{(\alpha} \tau_{\beta)} + 2 \tau_{(\alpha}\kappa_{\beta)\nu} h^{\mu\nu} \nonumber\\
        &\qquad - \frac{1}{2} {\geneNRT{T}}^\mu{}_{\alpha\beta} - h^{\mu \sigma} {\geneNRT{T}}^\nu{}_{\sigma (\alpha} b_{\beta)\nu}. \label{eq_mlmlm}
\end{align}
From this formula, we can define what is a Galilean contortion tensor $\hat K^\mu{}_{\alpha\beta}$, by assuming metricity, i.e.,~assuming ${\geneNR{\nabla}}_\alpha h^{\mu\nu} = 0$ and ${\geneNR{\nabla}}_\mu \tau_\nu = 0$:
\begin{align} 
    -\hat K^\mu{}_{\alpha\beta} &\coloneqq  \hat\Gamma^\mu{}_{\alpha\beta} - {\geneNR{\Gamma}}^\mu{}_{\alpha\beta} \quad \left({\rm with} \quad {\geneNR{\nabla}}_\alpha h^{\mu\nu} = 0, \ {\geneNR{\nabla}}_\mu \tau_{\nu} = 0\right) \label{eq_K_NR} \\
    &= h^{\mu\sigma}\left({\geneNR{\nabla}}_{(\alpha} b_{\beta)\sigma} - \frac{1}{2} {\geneNR{\nabla}}_{\sigma} b_{\alpha\beta} \right) + 2 \tau_{(\alpha}\kappa_{\beta)\nu} h^{\mu\nu} - \frac{1}{2} {\geneNRT{T}}^\mu{}_{\alpha\beta} - h^{\mu \sigma} {\geneNRT{T}}^\nu{}_{\sigma (\alpha} b_{\beta)\nu}. \nonumber 
\end{align}
We can also define the Galilean distorsion tensor $\hat L^\mu{}_{\alpha\beta}$, by assuming $\geneNR{\nabla}$ to be torsionless, i.e., $\geneNRT T^\mu{}_{\alpha\beta}=0$:
\begin{align}\label{eq_L_NR}
    -\hat L^\mu{}_{\alpha\beta} &\coloneqq  \hat\Gamma^\mu{}_{\alpha\beta} - {\geneNR{\Gamma}}^\mu{}_{\alpha\beta} \qquad \left({\rm with} \quad \geneNRT{T}^\mu{}_{\alpha\beta} = 0\right) \\
        &= h^{\mu\sigma}\left({\geneNR{\nabla}}_{(\alpha} b_{\beta)\sigma} - \frac{1}{2} {\geneNR{\nabla}}_{\sigma} b_{\alpha\beta} \right) + B^\mu {\geneNR{\nabla}}_{(\alpha} \tau_{\beta)} + 2 \tau_{(\alpha}\kappa_{\beta)\nu} h^{\mu\nu}. \nonumber
\end{align}

Note that, contrary to the Lorentzian case, the relation $\hat\Gamma^\mu{}_{\alpha\beta} - {\geneNR{\Gamma}}^\mu{}_{\alpha\beta} =  -\hat L^\mu{}_{\alpha\beta}  -\hat K^\mu{}_{\alpha\beta}$ does not hold, as the definitions of $\hat K^\mu{}_{\alpha\beta}$ and $\hat L^\mu{}_{\alpha\beta}$ are only valid if the other one is set to zero.
This means that the notion of contortion and distorsion in the Galilean case is ill-defined for a general affine connection. Another way of seeing this is from the relation $\tau_\rho \geneNR{T}^\rho{}_{\mu\nu}	= - 2 \hat \nabla_{[\mu}  \tau_{\nu]}$, which shows that the torsion $\geneNR{T}^\rho{}_{\mu\nu}$ and the non-metricity of $\tau_\mu$, i.e. $\hat \nabla_{\mu}  \tau_{\nu}$, are not independent.

\subsection{The non-relativistic limit}\label{sec_NR_limit}

In this section, we define the non-relativistic limit we will use to obtain STENC from the relativistic trinity. 
The limit begins with an expansion of the relativistic objects in terms of powers of $c$, the speed of light. As there is no absolute velocity in non-relativistic physics, one takes $c \rightarrow \infty$, which can be thought of as `flattening' the null cones at all spacetime points. Essentially, ``in the limit the cones are all tangent to a family of hypersurfaces, each of which represents ``space'' at a given ``time'', which corresponds to the standard Newtonian picture of spacetime'' \cite{1986_Malament}.
More precisely, the fundamental ansatz of the non-relativistic limit is to consider a Lorentzian metric admitting of a Taylor series in terms of $\lambda := 1/c^2$ given by the following \cite{1976_Kunzle}:
\begin{align}\label{expansionraised}
\accentset{\lambda}{g}^{\mu\nu} &\coloneqq h^{\mu\nu}+\lambda \, \tayl{1}{g}^{\,\mu\nu} + \lambda^2 \, \tayl{2}{g}^{\,\mu\nu} + \bigO{\lambda^3}, \\
\label{expansionlowered}
\accentset{\lambda}{g}_{\mu\nu} &\coloneqq -\frac{1}{\lambda}\tau_{\mu} \tau_{\nu} + \tayl{0}{g}_{\mu\nu} + \lambda \, \tayl{1}{g}_{\mu\nu} + \bigO{\lambda^2},
\end{align}
where $h^{\mu\nu}$ is a tensor whose kernel is 1-dimensional and $\tau_{\mu}$ is a 1-form. The leading orders of this Taylor series are degenerate, contrary to the full Lorentzian metric.\footnote{\citet{1992_Rendall} showed that there is not much other choice than this ansatz for the leading orders, but also that odd orders in $c$ should be considered. However, for the first few orders that we consider, these odd orders are purely gauge. Therefore, not considering them should not change the result of the paper.}\saut

The defining relation $\accentset{\lambda}{g}^{\mu\alpha}\accentset{\lambda}{g}_{\mu\beta} = \delta^\alpha_\beta$ for the Lorentzian metric implies the orthogonality relation $\tau_\mu h^{\mu\nu} = 0$, along with the following formulae for $\tayl{1}{g}^{\,\mu\nu}$ and $\tayl{0}{g}_{\mu\nu}$:
\begin{align}
    \tayl{1}{g}^{\,\mu\nu}  &= -B^\mu B^\nu + k^{\mu\nu}, \\
    \tayl{0}{g}_{\mu\nu}    &= b_{\mu\nu} - 2\tau_{\mu}\, \tau_{\nu} \phi,
\end{align}
where $k^{\mu\nu}\, \tau_{\mu} \coloneqq 0$, and $B^\mu$ and $b_{\mu\nu}$ are defined after Eq.~\eqref{eq_Galilean_connection}. Finally, $\phi$ is an arbitrary scalar (see e.g. \cite{1976_Kunzle, 2021_Vigneron} for further discussion related to these objects). Both $B^\mu$ and $\phi$ have some gauge freedom coming from the infinitesimal gauge freedom in the definition of the Taylor series (see Appendix~F in \citep{2024_Vigneron} for a detailed discussion).\saut


The non-relativistic expansion series defined in Eqs.~\eqref{expansionraised}--\eqref{expansionlowered} in orders of $\lambda$ imply that the standard Levi-Civita connection expands as $\accentset{\lambda}{\Gamma}^\alpha{}_{\mu \nu} = \tayl{-1}{\Gamma}^\alpha{}_{\mu \nu} + \tayl{0}{\Gamma}^\alpha{}_{\mu \nu} + \bigO{\lambda}$. However, it is only the zeroth order of the expansion of this connection that transforms as a connection, so it is only this order that can properly serve as a connection for the theories that emerge in this limit \cite{2017_Van-den-Bleeken}. Consequently, $\tayl{-1}{\Gamma}^\alpha{}_{\mu \nu}$ must vanish, which happens when we impose that $d\tau =0$ \cite{1976_Kunzle, 2023_Hartong_et_al}; or in other words, when $\tau$ is closed, which gives us a notion of absolute time inherent in standard Newtonian spacetime theories. The zeroth order of the expansion of the Levi-Civita connection then defines a (symmetric) Galilean connection $\hat{\nabla}_{\alpha}$ compatible with $h^{\mu\nu}$ and $\tau_\mu$, as introduced in Sec.~\ref{sec:nonrel}.\saut 

It should be noted that this is not the only way one can proceed. As detailed in \cite{2019_Hansen_et_al_b, 2020_Hansen_et_al}, one can take the non-relativistic limit of GR using a more general connection that does not force us to impose any conditions on $\tau$. When we do impose $d\tau = 0$, we recover familiar NC, which has been dubbed `Type I' Newton--Cartan theory in these papers, whereas relaxing this condition to $\tau \wedge \mathrm{d} \tau=0$ leads to `Type II' Newton--Cartan theories. Here we will adopt the condition that $\tau_{\mu}$ is closed, even exact, thereby staying within the realm of traditional Newtonian spacetime theories with a notion of absolute time. We proceed now to take the non-relativistic limit of STEGR, which will involve taking the non-relativistic limit of the distorsion tensor $L\indices{^\alpha_{\mu\nu}}$ that distinguishes this theory from GR.

\subsection{Distorsion and contortion in the non-relativistic limit}\label{sec_cont_dis_NR_limit}

\subsubsection{General formulae}

If the Galilean connection $\hat \Gamma^\mu{}_{\alpha\beta}$ derives from the non-relativistic limit (as defined in the previous section) of a Lorentzian connection $\accentset{\lambda}{\Gamma}^\mu{}_{\alpha\beta}$, assuming $\tau_{\mu}$ to be closed, then it has the form:
\begin{align}
 \hat\Gamma^\mu{}_{\alpha\beta} \coloneqq \tayl{0}{\Gamma}^\mu{}_{\alpha\beta} = 
        h^{\mu\sigma}\left(\partial_{(\alpha} b_{\beta)\sigma} - \frac{1}{2} \partial_{\sigma} b_{\alpha\beta} \right) + B^\mu \partial_{(\alpha} \tau_{\beta)} + \tau_{\alpha}\tau_\beta  h^{\mu\nu} \partial_\nu \phi. \label{eq_gene_diff_Gamma_NR}
\end{align}
Compared to a general Galilean connection~\eqref{eq_Galilean_connection}, the connection deriving from a non-relativistic limit has an exact Coriolis field, as $\kappa_{\mu\nu} = \tau_{[\mu} \partial_{\nu]} \phi$.\saut

Now the goal is to take the limit of the contortion $\accentset{\lambda}{K}^\mu{}_{\alpha\beta}$ [defined by~\eqref{eq_def_K_GR}], the distorsion $\accentset{\lambda}{L}^\mu{}_{\alpha\beta}$ [defined by~\eqref{eq_def_L_GR}], and the difference tensor ${\Gamma}^\mu{}_{\alpha\beta} - \gene{\Gamma}^\mu{}_{\alpha\beta}$.
As for the Levi-Civita connection, the general affine connection $\gene{\Gamma}^\mu{}_{\alpha\beta}$ should, {\it a priori}, have a Taylor series  as a function of~$\lambda$.
Let us assume that the leading order of this series is the zeroth order, i.e., $\accentset{\lambda}{\gene{\Gamma}}^\mu{}_{\alpha\beta} = \tayl{0}{\gene{\Gamma}}^\mu{}_{\alpha\beta} + \bigO{\lambda}$. We will discuss this hypothesis in Sec.~\ref{sec_Limit_flat_connection}. We obtain for the contortion
\begin{align}
    -\accentset{\lambda}{{K}}^\mu{}_{\alpha\beta} &= 
        \frac{1}{\lambda} \left[ h^{\sigma\mu} \tayl{0}{\geneT{T}}^{\,\gamma}{}_{\sigma(\alpha}\tau_{\beta)} \tau_\gamma \right] \nonumber\\
        &\qquad - \frac{1}{2} \tayl{0}{\geneT{T}}^{\,\mu}{}_{\alpha\beta} - h^{\mu \sigma}\tayl{0}{\geneT{T}}^{\,\gamma}{}_{\sigma (\alpha} \tayl{0}{g}_{\beta)\gamma}
        + \accentset{(1)}{g}^{\,\sigma\mu} \tayl{0}{\geneT{T}}^{\,\gamma}{}_{\sigma(\alpha} \tau_{\beta)} \tau_\gamma+ h^{\sigma\mu} \tayl{1}{\geneT{T}}^{\,\gamma}{}_{\sigma(\alpha} \tau_{\beta)} \tau_\gamma + \bigO{\lambda}, \label{eq_gene_K_Limit}
\end{align}
and for the distorsion
\begin{align}
    -\accentset{\lambda}{{L}}^\mu{}_{\alpha\beta} &= 
        \frac{1}{\lambda} \left[  h^{\sigma\mu}\left(\tau_\alpha \tayl{0}{\gene\nabla}_{[\sigma}\tau_{\beta]} + \tau_\beta \tayl{0}{\gene\nabla}_{[\sigma}\tau_{\alpha]}\right) \right] \nonumber\\
        &\qquad- h^{\mu\sigma}\tayl{1}{\geneT{T}}^{\,\gamma}{}_{\sigma(\alpha}\tau_{\beta)} \tau_\gamma + 
        \left(2\phi h^{\sigma\mu} + \accentset{(1)}{g}^{\,\sigma\mu}\right) \left(\tau_\alpha \tayl{0}{\gene\nabla}_{[\sigma}\tau_{\beta]} + \tau_\beta \tayl{0}{\gene\nabla}_{[\sigma}\tau_{\alpha]}\right) \nonumber \\
        &\qquad+ h^{\sigma\mu}\left(\tayl{0}{\gene\nabla}_{(\alpha} b_{\beta)\sigma} - \frac{1}{2} \tayl{0}{\gene\nabla}_\sigma b_{\alpha\beta}\right) + B^\mu \tayl{0}{\gene\nabla}_{(\alpha}\tau_{\beta)} + \tau_\alpha\tau_\beta h^{\mu\sigma}\partial_\sigma \phi + \mathcal{O}(\lambda), \label{eq_gene_L_Limit}
\end{align}
with $\tayl{0}{\gene{\nabla}} \coloneqq \partial + \tayl{0}{\gene{\Gamma}}$. These two formulae imply that the limit of the relativistic formula~\eqref{eq_gene_diff_Gamma} is
\begin{align}
    \accentset{\lambda}{\Gamma}^\mu{}_{\alpha\beta} - \accentset{\lambda}{\gene{\Gamma}}^\mu{}_{\alpha\beta} &=
        h^{\sigma\mu}\left(\tayl{0}{\gene\nabla}_{(\alpha} b_{\beta)\sigma} - \frac{1}{2} \tayl{0}{\gene\nabla}_\sigma b_{\alpha\beta}\right) + B^\gamma \tayl{0}{\gene\nabla}_{(\alpha}\tau_{\beta)} + \tau_\alpha\tau_\beta h^{\gamma\sigma}\partial_\sigma \phi \nonumber \\
        &\qquad- \frac{1}{2} \tayl{0}{\geneT{T}}^{\,\mu}{}_{\alpha\beta} - h^{\mu \sigma}\tayl{0}{\geneT{T}}^{\,\gamma}{}_{\sigma (\alpha} b_{\beta)\gamma} + \bigO{\lambda}, \label{eq_gene_diff_Gamma_Limit}
\end{align}
where we used the fact that $\tau_{\mu}$ is closed, implying $\tayl{0}{\gene\nabla}_{[\sigma}\tau_{\beta]} = -\frac{1}{2}\tayl{0}{\geneT{T}}^{\,\gamma}{}_{\alpha\beta} \, \tau_\gamma$.\saut

We see that, in general, the contortion and the distorsion tensors have negative orders in the limit. However, the difference tensor~\eqref{eq_gene_diff_Gamma_Limit} only has positive orders. Furthermore, this tensor does not depend on first orders of the torsion tensor contrary to the contortion and the distorsion. The formula~\eqref{eq_gene_diff_Gamma_Limit} is very similar to the general Galilean formula~\eqref{eq_mlmlm} when associating the zeroth order $\tayl{0}{\gene{\nabla}}$ with ${\geneNR{\nabla}}$, the only difference being the exactness of the Coriolis field.

\subsubsection{Non-relativistic limit with torsion and metricity}

We assume that $\accentset{\lambda}{\gene{\Gamma}}^{\mu}{}_{\alpha\beta}$ is metric. In the limit, this implies
\begin{align} \label{eq_NR_metricty_T}
    \tayl{0}{\gene{\nabla}}_\mu h^{\alpha\beta} = 0 \quad ; \quad  \tayl{0}{\gene{\nabla}}_\mu \tau_{\nu} = 0,
\end{align}
implying $\tayl{0}{\geneT{T}}^{\,\gamma}{}_{\alpha\beta} \, \tau_\gamma = 0$. Using $\accentset{\lambda}{L}^\mu{}_{\alpha\beta} = 0$, we also have
\begin{align}
    h^{\mu\sigma}\tayl{1}{\geneT{T}}^{\,\gamma}{}_{\sigma(\alpha} \tau_{\beta)} \tau_\gamma = h^{\sigma\mu}\left(\tayl{0}{\gene\nabla}_{(\alpha} b_{\beta)\sigma} - \frac{1}{2} \tayl{0}{\gene\nabla}_\sigma b_{\alpha\beta}\right) + \tau_\alpha\tau_\beta h^{\gamma\sigma}\partial_\sigma \phi.
\end{align}
Finally, the contortion tensor in the limit is
\begin{align}
    -\accentset{\lambda}{K}^\mu{}_{\alpha\beta} &= h^{\sigma\mu}\left(\tayl{0}{\gene\nabla}_{(\alpha} b_{\beta)\sigma} - \frac{1}{2} \tayl{0}{\gene\nabla}_\sigma b_{\alpha\beta}\right) + \tau_\alpha\tau_\beta h^{\gamma\sigma}\partial_\sigma \phi \nonumber \\
    &\qquad - \frac{1}{2} \tayl{0}{\geneT{T}}^{\,\mu}{}_{\alpha\beta} - h^{\mu \sigma}\tayl{0}{\geneT{T}}^{\,\gamma}{}_{\sigma (\alpha} b_{\beta)\gamma} + \bigO{\lambda}.
\end{align}
Compared to the general formula~\eqref{eq_K_NR} for Galilean contortion, the one obtained in the limit features an exact Coriolis field.
That form of the contortion tensor at zeroth order of the non-relativistic limit is in agreement with the one derived by \citet[][Eq.~(2.33a)]{2023_Schwartz}. His ``$\left(\tau_{(\alpha} f_{\beta)}{}^\mu\right)$'' is replaced by ``$\frac{1}{2} \tayl{0}{\geneT{T}}^{\,\mu}{}_{\alpha\beta} - h^{\sigma\mu}\left(\tayl{0}{\gene\nabla}_{(\alpha} b_{\beta)\sigma} - \frac{1}{2} \tayl{0}{\gene\nabla}_\sigma b_{\alpha\beta}\right) - \tau_\alpha\tau_\beta h^{\gamma\sigma}\partial_\sigma \phi$'' in our case, which has the same properties when projected along $h^{\mu\nu}$ and $\tau_{\mu}$. From the metricity relations~\eqref{eq_NR_metricty_T}, we can also show that
\begin{align}
    \tayl{0}{\geneT T}^{\, \mu}{}_{\alpha\beta} \tau_{\mu} = 0 \quad ; \quad \tayl{0}{\geneT T}^{\, (\alpha}{}_{\mu\nu} \, h^{\beta)\mu} h^{\nu \sigma} = 0.
\end{align}


\subsubsection{Non-relativistic limit without torsion and with non-metricity}

We assume that $\accentset{\lambda}{\gene{\Gamma}}^{\mu}{}_{\alpha\beta}$ is torsionless, which implies $\tayl{0}{\geneT{T}}^{\, \mu}{}_{\alpha\beta} = 0$ and $\tayl{1}{\geneT{T}}^{\, \mu}{}_{\alpha\beta} = 0$. Then the distorsion tensor becomes in the limit
\begin{align}\label{NRLimitDistortionTensor}
    -\accentset{\lambda}{\gene{L}}^\mu{}_{\alpha\beta} &= 
        h^{\sigma\mu}\left(\tayl{0}{\gene\nabla}_{(\alpha} b_{\beta)\sigma} - \frac{1}{2} \tayl{0}{\gene\nabla}_\sigma b_{\alpha\beta}\right) + B^\mu \tayl{0}{\gene\nabla}_{(\alpha}\tau_{\beta)} + \tau_\alpha\tau_\beta h^{\mu\sigma}\partial_\sigma \phi + \mathcal{O}(\lambda).
\end{align}
Compared to the general formula~\eqref{eq_L_NR} for a general Galilean distorsion tensor, the one obtained in the limit features an exact Coriolis field.

\subsubsection{Limit of a flat (reference) connection}
\label{sec_Limit_flat_connection}


The specificity of taking the non-relativistic limit of TEGR or STEGR compared to GR is that there is an additional field, namely $\gene\nabla$, whose behaviour in the limit needs to be understood. In the previous sections, the formulae derived and depending on~$\gene\nabla$ are valid regardless of the Riemann curvature of that connection. However, in the case of interest for this paper, the connection $\gene{\Gamma}$ is flat. Therefore, each order of its Riemann tensor must be zero.  Since we assumed that the leading order of the connection is the zeroth order, then that order is a connection whose Riemann tensor $\tayl{0}{\gene{R}}^{\,\mu}{}_{\alpha\beta\nu}$ is zero. In other words,  this implies that the zeroth order of the connection $\accentset{\lambda}{\gene \Gamma}^\mu{}_{\alpha\beta}$ is also flat. Therefore, the connection $\tayl{0}{\gene\nabla}$ appearing in the distorsion and contortion tensors in the non-relativistic limit is flat. This was expected of course, but not direct considering the fact  the reference connection has, in general, a Taylor series. This zeroth order will correspond to the flat affine connection present in the non-relativistic trinity, and will be denoted by $\geneNR\nabla$.

\begin{adjustwidth}{.5cm}{}
\begin{remark}
That results obtained in the limit hold only if $\accentset{\lambda}{\gene{\Gamma}}$ has positive orders. If we allow for negative orders, then the flatness condition $\accentset{\lambda}{\gene{R}}^{\,\mu}{}_{\alpha\beta\nu} = 0$ implies the constraint
\begin{align}
    \tayl{N}{\geneT{\Gamma}}^\alpha{}_{\mu \lambda} \tayl{N}{\geneT{\Gamma}}^\lambda{}_{\nu \beta} = \tayl{N}{\geneT{\Gamma}}^\alpha{}_{\nu \lambda} \tayl{N}{\geneT{\Gamma}}^\lambda{}_{\mu \beta},
\end{align}
where $N<0$ is the leading order, which is here, by definition, strictly negative. As for the Levi-Civita connection $\accentset{\lambda}{\Gamma}^\mu{}_{\alpha\beta}$, because only the zeroth order of the connection is not a tensor, all the other orders are tensors, then $\tayl{N}{\geneT{\Gamma}}$ is a tensor. Therefore, the above condition is a tensor equation.
We suspect that this constraint implies $\tayl{N}{\geneT{\Gamma}} = 0$, which would mean that the reference connection necessarily has positive orders once it is considered flat, hence justifying the hypothesis. But we have not been able to prove such a result.
\end{remark}
\end{adjustwidth}

\subsection{Non-relativistic limit of STEGR}\label{sec_GR_NR_limit}

In the case of taking the non-relativistic limit of TEGR, the prescription followed in \cite{2023_Schwartz} is the following one: (a) write the Einstein equation of GR in terms of the TEGR contortion (and associated torsion); (b) take the non-relativistic limit---constructed in the above way---of that equation. Following the same prescription for the non-relativistic limit of STEGR, we first recall the expression in Eq.~\eqref{eq_STEGR}, and take the non-relativistic limit. Taking the non-relativistic limit of the rhs gives \cite{1976_Kunzle}:
\begin{equation}
    \frac{8\pi G}{c^4}\left(T_{\mu\nu} - \frac{T^\alpha{}_\alpha}{2}g_{\mu\nu}\right) \rightarrow 4\pi G \rho \tau_{\mu}\tau_{\nu}.
\end{equation}
The validity of this formula requires the leading orders of the energy-momentum tensor $\accentset{\lambda}{T}^{\mu\nu}$ to be positive \citep{2024_Vigneron}. Cases where these orders are negative have been studied in \citep{2020_Hansen_et_al}, dubbed ``strong gravity''. In the present paper, we will not consider this possibility.\saut

We proceed now to take the limit of the lhs of the field equation~\eqref{eq_STEGR}, which is the Ricci tensor of the Levi-Civita connection expressed in terms of the distorsion and the non-metric connection. As mentioned earlier, when taking the non-relativistic limit of the connection, we will be seeking the zeroth order term in the expansion. This means that we will be searching for $\hat{L}\indices{^\alpha_{\mu \nu}} \coloneqq \tayl{0}{L}\indices{^{\alpha}_{\mu\nu}} = \tayl{0}{\gene{\Gamma}}\indices{^\alpha_{\mu \nu}} - \tayl{0}{{\Gamma}}\indices{^\alpha_{\mu \nu}} = \geneNR{\Gamma}\indices{^\alpha_{\mu \nu}} - \hat{\Gamma}\indices{^\alpha_{\mu \nu}}$, which is given by the leading order term in Eq.~\eqref{NRLimitDistortionTensor}. Consequently, the non-relativistic limit of the field Eq.~\eqref{eq_STEGR} of STEGR is:
\begin{align}\label{STNG field equation}
    -\geneNR {\nabla}_\alpha \hat{L}\indices{^\alpha_{\mu\nu}} + \geneNR{\nabla}_\mu \hat{L}\indices{^\alpha_{\nu\alpha}} 
    -\hat{L}\indices{^\alpha_{\mu \beta}} \hat{L}\indices{^\beta_{\alpha \nu}}+\hat{L}\indices{^\alpha_{\alpha \beta}} \hat{L}\indices{^\beta_{\mu \nu}} = 4\pi G \rho \tau_{\mu}\tau_{\nu},
\end{align}
where
\begin{align}\label{eq_erfzf}
\hat{L}^\alpha{ }_{\mu \nu} = - h^{\gamma \alpha}\left(\geneNR\nabla_{(\mu} b_{\nu)\gamma}  - \frac{1}{2}\geneNR\nabla_\gamma b_{\mu\nu}\right) -  B^\alpha \geneNR\nabla_{(\mu} \tau_{\nu)} - \tau_\mu \tau_\nu h^{\alpha \gamma} \partial_\gamma \phi.
\end{align}
This is the field equation of STENC.\saut

As it is formulated in Eq.~\eqref{eq_erfzf}, the non-relativistic distorsion tensor is not written explicitly as a function of a single non-metricity tensor as in the relativistic case~\eqref{eq_def_L_GR}. There are two reasons for this: first, we do not have a single metric like in GR, and second, this distorsion tensor does not depend only on these two metrics, but also on $B^\mu$ and~$b_{\mu\nu}$. This means that there are four different non-metricities that can naturally be defined in the STENC formulation:
\begin{align}\label{eq_def_NR_non_metricity}
    \hat Q_\mu{}^{\alpha\beta} \coloneqq \geneNR\nabla_\mu h^{\alpha\beta} \quad &; \quad \hat Q_{\mu\nu} \coloneqq \geneNR\nabla_\mu \tau_\nu \quad ; \quad 
    \hat Q_{\alpha\mu\nu} \coloneqq \geneNR\nabla_\alpha b_{\mu\nu} \quad ; \quad \hat Q_\mu{}^\nu \coloneqq \geneNR\nabla_\mu B^\nu.
\end{align}
We recall that since there is no duality between forms and vectors with Galilean structures, i.e. no way of raising and lowering indices, then $\hat Q_\mu{}^\nu$ and $\hat Q_{\mu\nu}$ (as well as $\hat Q_{\mu\alpha\beta}$ and $\hat Q_\mu{}^{\alpha\beta}$) have to be understood as different tensors.\saut

While the expression given above in Eq.~\eqref{eq_erfzf} for the distorsion tensor is the most straightforward one to write down from the non-relativistic limit, it would be desirable to be able to express the distorsion tensor mostly, or exclusively if possible, in  terms of the non-metricities encoded in the metric data $\tau_\mu$ and $h^{\mu\nu}$. It turns out that this is mostly possible and that one can equivalently write Eq.~\eqref{eq_erfzf} as (see Appendix~\ref{app_derivation} for a full derivation):\footnote{We are very grateful to Philip Schwartz for pointing this out to us. See \cite{Schwartz_2024aee} for further discussion.}
\begin{align}\label{true_distortion}
    \hat{L}^\alpha{ }_{\mu \nu} 
        &= - b_{\sigma(\mu}\hat Q_{\nu)}{}^{\sigma\alpha} + B^\alpha B^\sigma \tau_{(\mu} \hat Q_{\nu)\sigma} +\frac{1}{2} h^{\alpha \lambda} b_{\sigma\mu} b_{\nu \rho} \hat Q_{\lambda}{}^{\sigma\rho} + 2  \left[\tau_{(\mu} \geneNR\kappa_{\nu) \sigma} - \tau_{(\mu} \kappa_{\nu) \sigma}\right] h^{\sigma\alpha},
\end{align}
where $\kappa_{\mu\nu} = b_{\sigma[\nu} \hat\nabla_{\mu]} B^\sigma = \tau_{[\mu} \partial_{\nu]} \phi$ is the standard Coriolis field (in the non-relativistic limit) defined with respect to the Galilean connection, and $\geneNR\kappa_{\mu\nu} \coloneqq b_{\sigma[\nu} \geneNR\nabla_{\mu]} B^\sigma$ is 2-form defined similarly but with respect to the general connection $\geneNR\Gamma^\rho{}_{\mu\nu}$ (see Appendix~\ref{app_Coriolis} for more details). 
This expression mostly makes reference to the ``true'' non-metricities  $\hat Q_{\mu\nu}$ and $\hat Q_\mu{}^{\alpha\beta}$ defined in terms of the fixed temporal and spatial metrics that define the Galilean structure. But it also depends on the non-metricity of $B^\mu$, through the tensor $\geneNR\kappa_{\mu\nu}$, illustrating the fact that one cannot express the distorsion tensor solely as a function of the non-metricities of $\tau_\mu$ and $h^{\mu\nu}$. 
Equivalently, one cannot express the distorsion tensor solely as a function of the non-metricities of the (gauge-dependent) tensors $B^\mu$ and $b_{\mu\nu}$. From~\eqref{eq_erfzf}, we have the expression
\begin{align}\label{false_distortion}
\hat{L}^\alpha{ }_{\mu \nu} = -h^{\gamma \alpha}\left(\hat{Q}_{(\mu\nu)\gamma}  - \frac{1}{2}\hat{Q}_{\gamma\mu\nu}\right) - B^\alpha \hat{Q}_{(\mu\nu)} - \tau_\mu \tau_\nu h^{\alpha \gamma} \partial_\gamma \phi.
\end{align}
Compared to~\eqref{eq_erfzf}, the expressions~\eqref{true_distortion} and~\eqref{false_distortion} for the non-relativistic distorsion tensor are closer inline with its relativistic cousin~\eqref{eq_def_L_GR} explicitly written in terms of the Lorentzian non-metricity. However, for formulating STENC, the original expression~\eqref{eq_def_L_GR} has the advantage of explicitly showing where the gauge freedom relative to $B^\mu$ is present. Still, for the purpose of making specific choice of non-metricities (done in Sec.~\ref{sec_Choice}), expressions~\eqref{true_distortion} and~\eqref{false_distortion} will be of interest, especially for the recovery of standard Newtonian gravity in the Poisson equation formulation.

\subsection{The non-relativistic trinity}\label{sec_NRTrinity}

To summarise, then, the non-relativistic limit of the trinity of general relativity gives the three following equivalent systems of equation for non-relativistic gravitation:\saut

\noindent (i) the curved--torsionless--metric formulation (NC) is
\begin{align}
    \begin{dcases}
	   &\hat R_{\mu\nu} = 4\pi G \rho \tau_{\mu} \tau_{\nu}, \\
        &\hat\nabla_\mu T^{\mu\alpha} = 0, \\
        &\hat\Gamma^\mu{}_{\alpha\beta}  = h^{\gamma\mu}\left(\partial_{(\alpha} b_{\beta)\mu} - \frac{1}{2} \partial_\mu b_{\alpha\beta}\right) + B^\gamma \partial_{(\alpha}\tau_{\beta)} + \tau_\alpha\tau_\beta h^{\gamma\mu}\partial_\mu \phi, \\
    \end{dcases}
\end{align}


\noindent (ii) the flat--torsionful--metric formulation (TENC) is
\begin{align}\label{eq_TENC}
    \begin{dcases}
	   &-\geneNR\nabla_\alpha  \hat K^\alpha{}_{\mu\nu} + \geneNR\nabla_\mu \hat K^\alpha{}_{\nu\alpha} - \hat K^\alpha{}_{\mu\beta} \hat K^\beta{}_{\alpha\nu} + \hat K^\alpha{}_{\alpha\beta} \hat K^\beta{}_{\mu\nu} + \geneNRT T^\alpha{}_{\beta\mu} \hat K^\beta{}_{\alpha\nu} = 4\pi G \rho \tau_{\mu} \tau_{\nu}, \\
        &\geneNR\nabla_\mu T^{\mu\alpha} - 2T^{\sigma(\alpha}\hat K^{\mu)}{}_{\mu\sigma} = 0, \\
        &-\hat {K}^\mu{}_{\alpha\beta}= h^{\sigma\mu}\left(\geneNR\nabla_{(\alpha} b_{\beta)\sigma} - \frac{1}{2} \geneNR \nabla_\sigma b_{\alpha\beta}\right) + \tau_\alpha\tau_\beta h^{\gamma\sigma}\partial_\sigma \phi - \frac{1}{2} \geneNRT{T}^{\mu}{}_{\alpha\beta} - h^{\mu \sigma} \geneNRT{T}^{\gamma}{}_{\sigma (\alpha} b_{\beta)\gamma}, \\
        &{\rm with} \quad \geneNR{R}^{\mu}{}_{\alpha\beta\nu} = 0 \quad ; \quad \geneNR \nabla_\mu h^{\alpha\beta} = 0 \quad ; \quad \geneNR \nabla_\mu \tau_{\nu} = 0.
    \end{dcases}
\end{align}
\noindent (iii) the flat--torsionless--non-metric formulation (STENC) is
\begin{align}\label{eq_STENC}
    \begin{dcases}
        &-\geneNR\nabla_\alpha  \hat L^\alpha{}_{\mu\nu} + \geneNR\nabla_\mu  \hat L^\alpha{}_{\nu\alpha} -  \hat L^\alpha{}_{\mu\beta}  \hat L^\beta{}_{\alpha\nu} +  \hat L^\alpha{}_{\alpha\beta} \hat L^\beta{}_{\mu\nu} = 4\pi G \rho \tau_{\mu} \tau_{\nu}, \\
        &\geneNR\nabla_\mu T^{\mu\alpha} - 2T^{\sigma(\alpha}\hat L^{\mu)}{}_{\mu\sigma} = 0, \\
        &-\hat{L}^\mu{}_{\alpha\beta} = 
        h^{\sigma\mu}\left(\geneNR\nabla_{(\alpha} b_{\beta)\sigma} - \frac{1}{2} \geneNR\nabla_\sigma b_{\alpha\beta}\right) + B^\gamma \geneNR\nabla_{(\alpha}\tau_{\beta)} + \tau_\alpha\tau_\beta h^{\gamma\sigma}\partial_\sigma \phi, \\
        &{\rm with} \quad \geneNR R^{\mu}{}_{\alpha\beta\nu} = 0 \quad ; \quad \geneNRT T^\mu{}_{\alpha\beta} = 0.
    \end{dcases}
\end{align}
We recall that $T^{\mu\nu}$ is the energy-momentum tensor.\saut

There is a gauge freedom in terms of how to express the Galilean connection $\hat\Gamma^\mu{}_{\alpha\beta}$, contortion tensor $\hat K^\mu{}_{\alpha\beta}$ and distorsion tensor $\hat L^\mu{}_{\alpha\beta}$. That gauge is related to the observer 4-velocity~$B^\mu$ with respect to which $b_{\mu\nu}$ is defined, and in $\phi$. The origin of this freedom can be traced back to the gauge freedom in the Taylor series of the Lorentzian metric when considering the non-relativistic limit \citep[][Appendix~F.]{2024_Vigneron}. This type of gauge is also the one present in cosmological perturbation theory. The three field equations and their solution is of course independent on the choice of $B^\mu$.\saut

\begin{figure}[t]
    \centering
    \includegraphics[width=.8\textwidth]{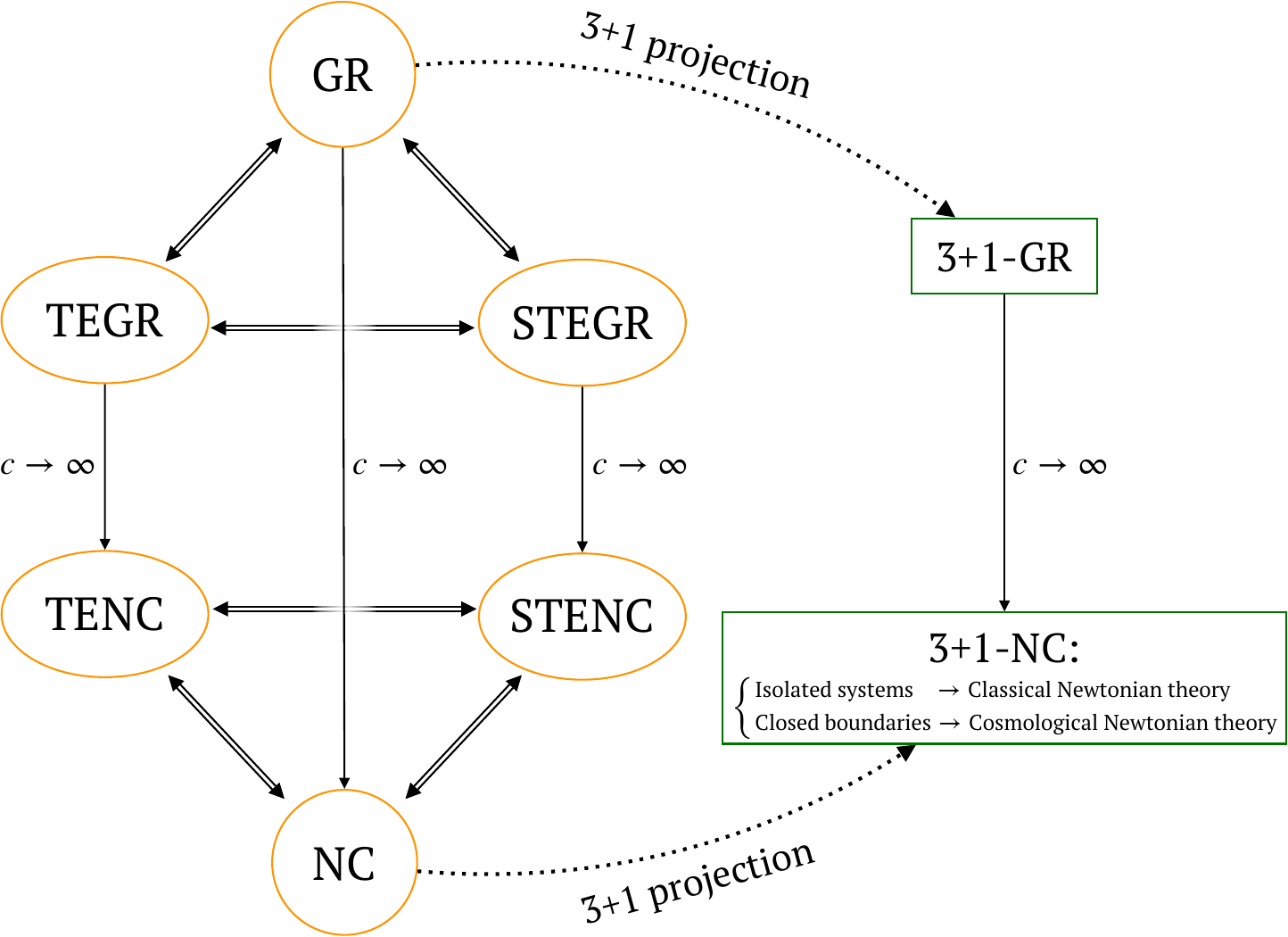}
    \caption{\label{fig_NR_Trinity_Final}
    The geometric trinity of gravity, its now-constructed non-relativistic counterpart, and the relations of both to the $3+1$ formulations of GR and NC, respectively.}
\end{figure}

In each case, the Coriolis field is exact, which implies the existence of vorticity-free observers and the absence of gravitomagnetism, two necessary features for the theory to be considered ``Newtonian''. Furthermore, since the first equations in each case are equivalent (considering the definition of $\geneNR\nabla$ along with the formulae for $\hat K^\mu{}_{\alpha\beta}$ and $\hat L^\mu{}_{\alpha\beta}$), and correspond to the Newton--Cartan equation, then each of these systems of equations are equivalent and lead to the classical Newtonian theory (for an isolated system) \citep{1963_Trautman} and to the cosmological Newtonian theory (if closed boundary conditions are considered) \citep{2021_Vigneron}, as explained in Sec.~\ref{sec_recovery}.
The first system has been known since \citet{1963_Trautman}, while the second was derived by \citet{2018_Read_et_al} and \citet{2023_Schwartz}. What is new in the present paper is the third system, defining the STENC formulation. Hence, this completes the construction of the non-relativistic trinity of gravity, as summarised in Figure~\ref{fig_NR_Trinity_Final}.\\

\begin{remark}{ 
Similarly to the relativistic case, we expect the last equations constraining the reference connection $\tilde\nabla$ in the systems~\eqref{eq_TENC} and \eqref{eq_STENC} to arise from Lagrange multipliers. However, at the non-relativistic level, no well-posed Lagrangian is known that leads to the fields equations of Newton-Cartan, and more generally to the three nodes of the non-relativistic trinity. Therefore, while, in theory, similar Lagrange multipliers should be present at the non-relativistic level compared to the relativistic one, the full Lagrangian is not known.}
\end{remark}

\subsection{Recovering the (classical) Newtonian theory}\label{sec_recovery}

Classical Newtonian theory is described by a system of equations defined on a 3-manifold. Therefore, recovering that theory from the Newton-Cartan formulation requires obtaining spatial equations from one of the Newton-Cartan systems of the non-relativistic trinity. The most general way of performing such a recovery is through a 3+1--projection where the 4-dimensional spacetime equations are projected along the time metric~$\tau_\mu$ (or along a timelike vector $V^\mu$) and the spatial metric $h^{\mu\nu}$. In the relativistic case, such a projection leads to the so-called 3+1--GR formulation, a way a describing the Einstein equation with equations defined on a 3-manifold, allowing for, in particular, numerical resolution of that equation~\citep{2012_GG}.\saut

When performing the 3+1--projection on the NC system, one does not directly obtain the classical Newtonian equations (in which, in particular, the Poisson equation is $\Delta_h \phi = 4\pi G \rho$, with $\Delta_h$ the spatial Laplacian), but similar equations featuring additional fields that depend on the choice of boundary conditions \citep{2021_Vigneron}.\footnote{This paper refers to `1+3' instead of `3+1' for the projection. In general relativity these two procedures differ depending on whether the time vector is vorticity free or not. However, in non-relativistic theories with an exact $\tau_\mu$, 3+1 and 1+3 projections become degenerate. We choose to use the `3+1' denomination in the present paper.} Only when the physical fields are considered integrable and the space infinite, in other words choosing an isolated system, then the 3+1-projection of the NC system leads to the classical Newtonian theory. However, if closed boundary conditions are chosen, which is more representative of a cosmological setup, then, instead, the cosmological Newtonian theory is retrieved, in which expansion is present. This was first studied by \citep{1997_Ruede_et_al} in the homogeneous case, and fully derived in  the general (inhomogeneous) case by \citep{2021_Vigneron}; see \citep{Wallace2020-WALFAE-5} for related foundational discussions.\saut

The fact that the NC formulation naturally features expansion is usually missed in references studying the recovery of (classical) Newtonian theory. The main reason is that the spatial metric is, in general, assumed to be $\delta_{ij}$ in some coordinate systems, i.e. independent of time, which imposes the vanishing of expansion. In other references \citep[e.g.,][]{2023_Schwartz}, a covariant condition is taken by assuming the existence of observers whose 4-velocity $u^\mu$ satisfies $h^{\mu(\alpha}\hat\nabla_\mu u^{\beta)} = 0$. This is exactly the covariant condition for vanishing expansion in NC, as shown in \citep{2021_Vigneron} (see also Sec.~\ref{sec_Observer}).\saut

Now, with regard to the recovery process from STENC, since that system of equations is equivalent to the NC formulation, the same recovery is therefore possible. This means that STENC leads to either the classical or the cosmological Newtonian theory, depending on the choice of boundary conditions.

\section{Discussion}\label{sec_discussion}

Having now constructed the non-relativistic geometric trinity of gravity, in this section we  discuss (i) observers in the non-relativistic trinity (Sec.~\ref{sec_Observer}), (ii) the special case of Weylian (i.e.,~pure trace) non-relativistic non-metricity (Sec.~\ref{sec_Weyl}), and (iii) different possible choices for non-relativistic non-metricity in relation to the Poisson equation (Sec.~\ref{sec_Choice}).

\subsection{Observers in the non-relativistic trinity}\label{sec_Observer}

In non-relativistic theories, a (timelike) observer is defined by a 4-velocity $u^\mu$ unit with respect to the time metric, i.e. $u^\mu \tau_\mu = 1$. Two types of observers are of major importance: geodesic observers and inertial observers. The former are related to the equivalence principle while the latter to the concept of inertial frames of Newtonian theory as well as the gravitational field: 

\begin{enumerate}[label=(\roman*)] 
\item {\it Geodesic observers:} As for GR, in non-relativistic gravitation an observer subject only to gravity follows the geodesics (related to the Galilean connection) of spacetime. In the trinity, this translates into the following three equations:
\begin{align}
    {\rm NC:}& \quad u^\mu \hat\nabla_\mu u^\alpha = 0, \\ 
    {\rm TENC:}& \quad u^\mu\geneNR\nabla_\mu u^\alpha = u^\mu u^\nu \hat K^\alpha{}_{\mu\nu}, \\
    {\rm STENC:}& \quad u^\mu\geneNR\nabla_\mu u^\alpha = u^\mu u^\nu \hat L^\alpha{}_{\mu\nu}.
\end{align}
These three equations are all equivalent. However, for TENC and STENC, we can change the point of view and say that, from (49) and (50), particles are deviated from the geodesics of the reference flat connection by a force coming, respectively, from the Galilean contorsion and the Galilean distorsion. Such an interpretation can likewise be made in the relativistic case.


\item {\it Inertial observers:} As shown in \citep{2021_Vigneron}, the NC equation implies the existence of an inertial observer (called Galilean observer in that paper) which is defined by a 4-velocity $G^\mu$ with the following constraints:\footnote{In terms of a (unique) scalar-vector-tensor decomposition of symmetric tensors, the first condition means that the expansion tensor $h^{\mu(\alpha} \hat \nabla_\mu G^{\beta)}$ of $G^\mu$ has no gradient part \citep{2021_Vigneron}.}
\begin{align}
    h^{\mu(\alpha} \hat \nabla_\mu G^{\beta)} &=  \chi h^{\alpha\beta} + \Xi^{\alpha\beta}, \\
    h^{\mu[\alpha} \hat \nabla_\mu G^{\beta]} &= 0,
\end{align}
where $\chi$ is a scalar field (imposed to be a spatial constant by the Newton--Cartan equation), representing global expansion, and $\Xi^{\mu\nu}$ is a traceless-harmonic spatial tensor (with respect to the spatial metric $h^{\mu\nu}$) representing anisotropic expansion. The value of the expansion fields $\chi$ and $\Xi^{\mu\nu}$ depend on boundary conditions. For an isolated system, where $\chi = 0$ and $\Xi^{\mu\nu} = 0$ (i.e. no expansion), the first condition implies that in a coordinate system adapted to the $\tau$-foliation and such that $\partial_t^\mu = G^\mu$, the spatial metric components are independent of time. If expansion is present with $\chi\not=0$ or $\Xi^{\mu\nu}\not=0$, one or several scale factors appear.
The second condition means that $G^\mu$ is vorticity free.
\end{enumerate}

The gravitational field $g^\mu$ in the Newton--Cartan formulation is elegantly defined as the opposite of the 4-acceleration of an inertial observer:
\begin{align}
    g^\alpha \coloneqq -G^\mu \hat\nabla_\mu G^\alpha. \label{eq_def_g}
\end{align}
From the Newton--Cartan equation and the exactness of the Coriolis field, we can show that $g^\mu$ is vorticity free (i.e. no gravitomagnetism), and that its potential, i.e. $g^\mu = h^{\mu\rho}\partial_\rho \Phi$, is the scalar field $\Phi$ entering into the Poisson equation \citep{2021_Vigneron}. These properties also follow from the definition of an inertial observer.
In the language of the non-relativistic trinity, the gravitational field can be written in the following forms:
\begin{align}\label{eq_g_SNGT}
    {\rm TENC:} &\quad g^{\alpha} = -G^\mu \geneNR\nabla_\mu G^\alpha + G^\mu G^\nu \hat K^\alpha{}_{\mu\nu}, \\
    {\rm STENC:} &\quad g^{\alpha} = -G^\mu \geneNR\nabla_\mu G^\alpha + G^\mu G^\nu \hat L^\alpha{}_{\mu\nu}.
\end{align}
\begin{adjustwidth}{.5cm}{}
\begin{remark}
    We suspect that in both cases the gauge freedom in the definition of $\geneNR\nabla$ allows us to assume $G^\mu \geneNR\nabla_\mu G^\alpha = 0$, with the elegant interpretation that inertial observers are not accelerating with respect to the flat connection. However, we were not able to prove this property.
\end{remark}
\end{adjustwidth}

\subsection{The Weylian special case}\label{sec_Weyl}

As we have been concerned with theories that exhibit general non-metricity, we consider now the special case of `Weylian' non-metricity, of the form
\begin{equation}\label{Weylcondition}
    Q_{\alpha \mu\nu} = \sigma_\alpha g_{\mu\nu},
\end{equation}
where $\sigma$ is a 1-form. In other words, we consider now the case in which the non-metricity is `pure trace'. This form of non-metricity is of both historical and modern interest.  Historically, Weyl famously generalized Riemannian geometry by relaxing the metric compatibility condition in a failed attempt to unify gravity and electromagnetism~\cite{1997_ORaifeartaigh}. While this particular attempt was unsuccessful, Weyl geometries and Weyl-inspired gravitational theories themselves are still an active field of research (see, e.g., \cite{Banados_2024, Ferreira_2016wem, Blas_2011ac, Klemm:2020mfp} for some recent work related to Weyl-inspired gravity). These theories are particularly interesting in how they manifest scale-invariance as they are invariant under the well-known `Weyl transformations'~\cite{2012_Romero, 2018_Wheeler}:
\begin{equation}\label{eq:WeylSym}
    \begin{aligned}
        g^{\mu\nu} &\mapsto e^{-f} g^{\mu\nu}, \\
        \sigma_\alpha &\mapsto \sigma_\alpha - \bar{\nabla}_\alpha f.
    \end{aligned}
\end{equation}
Furthermore, Weyl geometries have been attracting renewed interest following recent foundational work that demonstrates that they do not, in fact, suffer from the second clock effect once one properly models matter fields in such a theory \cite{Hobson_2020doi, Hobson_Lasenby_2022}.

Most relevantly to us, \cite{2012_Romero} shows how to formulate GR in the language of Weyl geometry by adopting the condition in Eq.~\eqref{Weylcondition}. This leads to a distorsion tensor given by:
\begin{equation}
L\indices{^\alpha_{\mu \nu}}=-\sigma_{(\mu} \delta_{\nu)}^\alpha+\frac{1}{2} \sigma^\alpha g_{\mu \nu}.
\end{equation}
In the following, we proceed to take the non-relativistic limit of this ansatz in the same way as before, using the Taylor expansion defined in Sec.~\ref{sec:limit}. First, we must consider the Taylor expansion of  $\sigma$:
\begin{align}
\accentset{\lambda}{\sigma}_\mu=\tayl{0}{\sigma}_\mu+\lambda \tayl{1}{\sigma}_\mu+\bigO{\lambda^2}.
\end{align}
The non-metricity in this Weylian case then becomes
\begin{align}
\geneNR{\nabla}_\alpha h^{\mu \nu}=-\tayl{0}{\sigma}_\alpha h^{\mu \nu} \quad ; \quad \geneNR{\nabla}_\mu \tau_\nu=\frac{1}{2} \tayl{0}{\sigma}_\mu \tau_\nu,
\end{align}
where, again, we denote $\tayl{0}{\gene\nabla}$ by $\geneNR\nabla$ which is flat and torsionless. (Note that this is consistent with \citep{2020_Dewar_et_al}.) We see that in the limit the Weylian form in the ansatz~\eqref{eq:WeylSym} remains present for $\geneNR\nabla_\mu h^{\alpha\beta}$ and $\geneNR\nabla_\mu \tau_\nu$ as these two derivatives are both proportional to the 1-form $\tayl{0}{\sigma}_\mu$ obtained in the limit. This suggests that conformal gravity theories should have a non-relativistic limit preserving the conformal symmetry, but it is transfered to the spatial and the temporal~metrics.\saut

Regarding the distorsion tensor, as before, when taking the non-relativistic limit, we are searching for 
$\hat{L}\indices{^\alpha_{\mu \nu}} \coloneqq \tayl{0}{L}\indices{^{\alpha}_{\mu\nu}} = \tayl{0}{\gene\Gamma}\indices{^\alpha_{\mu \nu}} - \hat{\Gamma}\indices{^\alpha_{\mu \nu}} = \geneNR{\Gamma}\indices{^\alpha_{\mu \nu}} - \hat{\Gamma}\indices{^\alpha_{\mu \nu}}$ while assuming $\gene\nabla$ to be torsionless, but with the `Weylian' versions of the non-metric objects. 
One obtains a {(-1)-order} term $\tayl{-1}{L}\indices{^\alpha_{\mu \nu}}=-\frac{1}{2} h^{\alpha \beta} \, \tayl{0}{\sigma}_\beta \tau_\mu \tau_\nu$ which has to vanish to obtain a Galilean connection in the limit. This gives us $h^{\alpha \beta} \, \tayl{0}{\sigma}_\beta=0$. In other words, ${\sigma}_\mu$ 
is proportional to the time metric, i.e., ${\sigma}_\mu = \psi \, \tau_\mu$, where $\psi$ is an arbitrary scalar. This implies
\begin{align}
    \tayl{0}{L}\indices{^\alpha_{\mu \nu}}
        &=-\tayl{0}{\sigma}_{(\mu} \delta_{\nu)}^\alpha-\frac{1}{2}\left(\tayl{1}{g}^{\,\alpha \beta}\,\tayl{0}{\sigma}_{\beta}+h^{\alpha \beta}\,\tayl{1}{\sigma}_{\beta}\right) \tau_\mu \tau_\nu \nonumber \\
        &= -\psi \, \tau_{(\mu} \delta^\alpha_{\nu)} + \frac{1}{2}\left(\psi B^\alpha - h^{\alpha \beta}\,\tayl{1}{\sigma}_{\beta}\right) \tau_\mu \tau_\nu.
\end{align}    
Using this result and $\geneNR{\nabla}_{[\mu}\tayl{0}{\sigma}_{\nu]} = 0$ (that $\sigma$ is closed is implied by the non-metricity condition), the field equations under the Weylian assumption for non-metricity are given by
\begin{equation}
-\geneNR{\nabla}_\mu \tayl{0}{\sigma}_{\nu} - \frac{1}{2}\tayl{0}{\sigma}_{\mu}\tayl{0}{\sigma}_{\nu} + \frac{1}{2} \geneNR{\nabla}_\alpha\left[\tayl{1}{g}^{\,\alpha \beta}\,\tayl{0}{\sigma}_{\beta}+h^{\alpha \beta}\,\tayl{1}{\sigma}_{\beta}\right] \tau_\mu \tau_\nu = 4 \pi G \rho \tau_\mu \tau_\nu.
\end{equation}

\subsection{Choice of non-metricity and the Poisson equation}\label{sec_Choice}

In this section, we show that there are two mirror ansatzes that can be imposed on the non-relativistic non-metricity tensors defined in Eq.~\eqref{eq_def_NR_non_metricity}, such that the field equation of STENC becomes explicitly the Poisson equation written in terms of a spacetime connection. The two conditions are:
\begin{enumerate}[label = (\roman*)]
    \item  $\hat Q_{\mu\alpha\beta} = 0$ and $\hat Q_{\mu}{}^\nu = 0$,
    \item $\hat Q_\mu{}^{\alpha\beta} = 0$ and $\hat Q_{\mu\nu} = 0$.
\end{enumerate}
    Both of these approaches restrict the possible physical solutions as spatial expansion is set to zero with $\hat Q_\mu{}^{\alpha\beta} = 0$ or $\hat Q_{\mu\alpha\beta} = 0$, meaning that only an isolated system can be considered. The goal of these ansatzes is mainly to illustrate how the physical degrees of freedom can be shifted between the different non-metricities.\saut

With the first choice, the distorsion tensor from Eq.~\eqref{false_distortion} becomes
\begin{align}\label{eq_jkjk}
    \hat{L}^\alpha{ }_{\mu \nu} = - B^\alpha \geneNR\nabla_{(\mu} \tau_{\nu)} -\tau_\mu \tau_\nu h^{\alpha \gamma} \partial_\gamma \phi,
\end{align}
which implies that $\hat{L}^\alpha{ }_{\alpha \mu} = B^\alpha \hat{Q}_{(\alpha\mu)} = \geneNR\nabla_\mu\left(B^\alpha \tau_\alpha\right) - \hat Q_{\mu}{}^\alpha \tau_\alpha = 0$. Consequently, the field equation simplifies to:
\begin{equation}
    -\geneNR{\nabla}_\alpha \hat{L}\indices{^\alpha_{\mu\nu}}
    -\hat{L}\indices{^\alpha_{\mu \beta}} \hat{L}\indices{^\beta_{\alpha \nu}} =
     4\pi G \rho \, \tau_{\mu}\tau_{\nu}.
\end{equation}
Carrying through the calculations, we have that
\begin{align}
    \begin{aligned}
     -\geneNR{\nabla}_\alpha \hat{L}\indices{^\alpha_{\mu\nu}} &= 2 \hat{Q}_{\alpha\mu} \tau_\nu h^{\alpha \gamma}\geneNR{\nabla}_\gamma \phi + \tau_\mu \tau_\nu \geneNR{\nabla}_\alpha \left( h^{\alpha\gamma}\geneNR{\nabla}_\gamma \phi\right), \\
     -\hat{L}\indices{^\alpha_{\mu \beta}} \hat{L}\indices{^\beta_{\alpha \nu}} &= -2 \hat{Q}_{\alpha\mu} \tau_\nu h^{\alpha \gamma}\geneNR{\nabla}_\gamma \phi.
    \end{aligned}
\end{align}
The field equation of STENC then takes the form
\begin{align}\label{eq_Poisson_Gbar_bis}
   \tau_\mu \tau_\nu \geneNR{\nabla}_\alpha \left( h^{\alpha\gamma}\geneNR{\nabla}_\gamma \phi\right) = 4\pi G \rho \tau_{\mu}\tau_{\nu},
\end{align}
which is the Poisson equation expressed in terms of a curvature-free, torsion-free  spacetime connection that also possesses non-metricity (with respect to $\tau_\mu$ and $h^{\mu\nu}$), related by $\geneNR{\nabla}_\alpha h^{\mu \nu}= 2 B^{(\mu} \hat{Q}_{\alpha\gamma}h^{\nu)\gamma}$. A similar equation was obtained in \citep[Eq.~31]{2018_Read_et_al} and \citep[Eq.~4.20]{2023_Schwartz} in the torsional case. These papers fixed both the expansion and spatial torsion to vanish, and obtained the `standard' formulation of Newtonian gravity in terms of a torsional connection rather that the non-metric connection here. See e.g. \cite{2023_March_et_al} for further discussion of these versions of Newtonian gravity that result from such geometric considerations.\saut

In the second choice, the distorsion tensor takes the form
\begin{align}\label{eq_jkjk_bis}
    \hat{L}^\alpha{ }_{\mu \nu} = 2\tau_{(\mu} \geneNR\kappa_{\nu)\sigma} h^{\sigma\alpha} -\tau_\mu \tau_\nu h^{\alpha \gamma} \partial_\gamma \phi,
\end{align}
implying that $\hat L^\alpha{}_{\alpha\mu} = 0$, leading the the field equation
\begin{align}\label{eq_Poisson_Gbar}
   \tau_\mu \tau_\nu h^{\alpha\beta}\geneNR{\nabla}_\alpha \geneNR{\nabla}_\beta \phi - 2h^{\sigma\gamma}\tau_{(\mu} \geneNR\nabla_\sigma \geneNR\kappa_{\nu)\gamma} = 4\pi G \rho \tau_{\mu}\tau_{\nu}.
\end{align}
Contrary to the first case, in the second case, all the gauge degrees of freedom have not been fixed, illustrated by the fact that the $B^\mu$ dependent term $\geneNR\kappa_{\mu\nu}$ remains present in the field equation. Consequently, the scalar $\phi$ in that equation is not yet the gravitational potential, which is a gauge invariant quantity. A natural gauge fixing is to assume $B^\mu = G^\mu$, i.e., the gauge dependent 4-velocity refers to an inertial observer, which is a gauge choice that can always be taken \citep{2021_Vigneron, 2024_Vigneron}. This implies that the gauge dependent scalar $\phi$ becomes the gravitational potential $\Phi$ present in the gravitational field $g^\mu = h^{\mu\nu} \partial_\nu \Phi$ that we defined in Sec.~\ref{sec_Observer}, and that $\geneNR\kappa_{\mu\nu} = 0$. We then again retrieve the Poisson equation expressed in terms of a curvature-free, torsion-free  spacetime connection.\saut


There is nevertheless a major conceptual difference between the two approaches, as with the second one, the connection $\geneNR\nabla$ is a Galilean connection, i.e. it is compatible with the space and time metrics. In this sense, arguably we lose to some degree the philosophy of the STENC formulation, where the goal was to have two symmetric connections: one Galilean and one non-Galilean.

\subsection{Gravitational stress-energy}

The field equations for TEGR and STEGR [Eqs.~\eqref{eq_TEGR} and \eqref{eq_STEGR}], and for TENC and STENC [Eqs.~\eqref{eq_TENC} and \eqref{eq_STENC}] are closely related to covariant conservation laws, of the form $\partial_\mu \left( T_{\mu\nu} + t_{\mu\nu} \right) = 0$, where $T_{\mu\nu}$ is the stress-energy tensor and $t_{\mu\nu}$ represents gravitational stress-energy. In GR, $t_{\mu\nu}$ is a pseudotensor and suffers from various degeneracies; however, it appears to be tensorial in TEGR \cite{2012_Aldrovandi_et_al} and STEGR \cite{1940_Rosen_a}. Still in these two cases, the gravitational stress-energy tensor depends upon the gauge choice on $\bar\Gamma$. This is related to the fact that, in GR, there is not a unique way of defining conserved quantities such as energy or mass (see also e.g.\ \cite{PittsEinstein, PittsEnergy} for further discussions). 

Given these interpretative difficulties, in addition to the facts that (i) the non-relativistic versions of such objects often seem to vanish identically (see \cite{DurrRead}) and (ii) to make rigorous sense of such conservation laws one needs to recourse to Noether's second theorem (see \cite{DeHaro} for a recent review), which is unavailable in the non-relativistic case given the absence of a suitable action, we set aside this approach in the present work, while acknowledging that it would be valuable and interesting to explore the matter further in future research. In our view, the right way to proceed regarding these issues will again be to avail oneself of `Type II' NC, in which an action principle (and, hence, the resources of Noether's theorems etc.)\ are available.

\section{Conclusion}\label{sec:close}

In this article, we have taken the non-relativistic limit of the symmetric teleparallel equivalent formulation of general relativity (STEGR), and have obtained the symmetric teleparallel equivalent formulation of Newton--Cartan theory (STENC). Just as in the relativistic case, the gravitational degrees of freedom in STENC are in the non-metricity; thereby, we have triangulated a non-metric alternative theory to Newton--Cartan theory (NC) and its equivalent teleparallel version (TENC). We have therefore completed a non-relativistic geometric trinity for gravity---this also makes good on a question raised in \cite{2023_Schwartz} as to what one would obtain on taking the non-relativistic limit of STEGR.\saut

In completing this work, we cast new light and understanding upon the relationship between the non-relativistic limit and geometrical reformulations of spacetime theories, as well as come to understand better the geography of the `space of spacetime theories' more generally (cf.~\cite{1973_Thorne_et_al, 2017_Lehmkuhl_et_al}). Moreover, the existence of the non-relativistic geometric trinity need not be a mere theoretical or philosophical curiosity: it is already known in the case of the relativistic geometric trinity that different vertices of the trinity are more or less apt to represent different physical scenarios (e.g., black hole boundaries---see \cite{2017_Oshita_et_al}); in principle, we expect the same to be true in the non-relativistic case, although we will leave such explorations for future work. There are many future prospects to the present work.
\begin{enumerate}[label=(\roman*)]
    \item Recently, in \cite{2019_Hansen_et_al_b}, a novel version of NC (so-called `Type II NC') has been constructed by taking a more careful and systematic $1/c$ expansion of the GR dynamics. Type II NC has revealed several novel features of non-relativistic gravitational theories, including that non-relativistic theories can account for many of the strong gravitational effects previously believed to belong to relativistic theories and can also reproduce much of the solution space of GR \cite{2019_Hansen_et_al, 2020_Hansen_et_al}. This raises the question: what would be the `Type II' equivalents of TENC and STENC? Indeed, finding such theories would have conceptual payoff, for in this article we have demonstrated equivalence of the three vertices of the non-relativistic geometric trinity only at the level of equations of motion, whereas the equivalence of the relativistic geometric trinity can---as we have seen---be demonstrated at the level of the action. However, action principles for the `Type I' theories provably do not exist \cite{2020_Hansen_et_al}; not so for `Type II' theories (and, indeed, an action for Type II NC is explicitly constructed in \cite{2020_Hansen_et_al}); therefore, to construct a non-relativistic trinity using action principles (as in the relativistic case), one would have to construct and work with the `Type II' theories.
    \item It has very recently been shown that there exists an `extended' geometric trinity between  $f(R)$, $f(T, B)$ and $f(Q , \tilde{B})$ theories (for boundary terms $B$ and $\tilde{B}$) \cite{2023_Capozziello_et_al, Erdmenger:2023hne}---does a similar extension of the non-relativistic geometric trinity exist? (For further discussion on this issue in the relativistic case, see \cite{2021_Bohmer_et_al, 2023_Bohmer_et_al}.).
    \item It has also been suggested that these non-relativistic theories can be used to develop coordinate free post-Newtonian expansions of TEGR and STEGR (as well as other theories in the broader metric-affine gravity framework) \cite{Schwartz_2024aee}, which could be especially useful for theorists working within the theory space of $f(T)$ and $f(Q)$ gravity.
    \item Given that both standard and torsional Newton-Cartan theory have been used to develop non-relativistic string theory and gain insights into string theory and holography \cite{Bergshoeff_2019pij}, we are hopeful that developing non-relativistic versions of non-metric geometries as we have done here will pave the way for further insights---especially as it relates to exploring fundamental aspects of $f(Q)$ and Weyl gravity theories of current interest and given that there is already interest in Weyl geometries within holography \cite{Ciambelli:2019bzz}.
    \item It is possible to understand different nodes of the relativistic geometric trinity as different gauge theories of gravity: GR can be understood as a gauge theory of the Lorentz transformations; TEGR as a gauge theory of the translations; and STEGR as a gauge theory of shear/scale (and---as mentioned above---also the translations) \cite{1995_Hehl_et_al}. This invites the question: can the nodes of the non-relativistic trinity also be understood as gauge theories of gravity in a similar manner? We are optimistic about the prospects for an affirmative answer here 
    but, again, we will leave a full study of this question for future work.
    \item And finally, we saw earlier (recall Sec.~\ref{sec:nonrel}) that introducing the geometric concept of torsion into Newton-Cartan theory proved to be incredibly fruitful for modelling condensed matter systems. Likewise, there are other emergent systems for which non-metricity seems capture and/or appropriately model phenomena for which curvature alone is unsuitable. For example, consider nonlinear elastic solids in solid state physics, where topological defects induce an effective geometry that modifies local descriptions of length \cite{Latorre_2017, Ghilencea_2022lcl}. From this description alone, it should not be surprising that non-metricity has proved to be a useful geometric concept in these contexts (see, e.g., \cite{Roychowdhury_2017, Yavari_2012}). We believe that developing the mathematical framework for non-metricity in a non-relativistic context as we have done here will have important practical utility for modeling phenomena in solid state physics because the non-relativistic framework will be more appropriate these systems than the more well-known relativistic framework---just as non-relativistic geometry has proven to be particularly useful for condensed matter systems.
\end{enumerate}

\section*{Acknowledgements}
We are very grateful to Jelle Hartong, Nic Teh, and Philip Schwartz for helpful discussions. W.W.~acknowledges support from St. Cross College, Oxford. J.R.~acknowledges the support of the Leverhulme Trust. Q.V.~acknowledges the support of the Centre of Excellence in Astrophysics and Astrochemistry of Nicolaus Copernicus University in Toru\'n, and of the Polish National Science Centre under Grant SONATINA No. 2022/44/C/ST9/00078.\saut


\appendix

\section{Freedom on the Coriolis field}\label{app_Coriolis}

\subsection{Non-uniqueness}
For a given Galilean connection $\hat \Gamma^\alpha{}_{\mu\nu}$ compatible with $\tau_\mu$ and $h^{\mu\nu}$, the Coriolis field $\kappa_{\mu\nu}$ is not uniquely defined. It depends on the choice of vector $B^\mu$ present in the formula~\eqref{eq_Galilean_connection}. To see this, let us consider a timelike vector $G^\mu = B^\mu - v^\mu$ where $v^\mu \tau_\mu \coloneqq 0$. The projector orthogonal to $B^\mu$, which we denoted $b_{\mu\nu}$, can be expressed a function of the projector orthogonal to $G^\mu$, denoted by ${^G}b_{\mu\nu}$, as follows:
\begin{align}
    b_{\mu\nu} = {^G}b_{\mu\nu} - 2 v^\sigma \, {^G}b_{\sigma(\mu} \tau_{\nu)} + v^\sigma v^\rho \, {^G}b_{\sigma\rho} \, \tau_\mu \tau_\nu.
\end{align}
Let us rewrite the Galilean connection~\eqref{eq_Galilean_connection} as a function of $G^\mu$ and its projector. First, we consider the connections ${^G}\Gamma^\alpha{}_{\mu\nu}$ and ${^B}\Gamma^\alpha{}_{\mu\nu}$ defined as:
\begin{align}
    {^G}\Gamma^\alpha{}_{\mu\nu} \coloneqq h^{\mu\sigma}\left(\partial_{(\alpha} \, {^G}b_{\beta)\sigma} - \frac{1}{2} \partial_{\sigma} \, {^G}b_{\alpha\beta} \right) + G^\mu \partial_{(\alpha} \tau_{\beta)},\\
    {^B}\Gamma^\alpha{}_{\mu\nu} \coloneqq h^{\mu\sigma}\left(\partial_{(\alpha} \, {^B}b_{\beta)\sigma} - \frac{1}{2} \partial_{\sigma} \, {^B}b_{\alpha\beta} \right) + B^\mu \partial_{(\alpha} \tau_{\beta)},
\end{align}
where we have $\hat\Gamma^\alpha{}_{\mu\nu} = {^B}\Gamma^\alpha{}_{\mu\nu} + 2\tau_{(\mu} \kappa_{\nu)\sigma}h^{\sigma\alpha}$, i.e. the connection ${^B}\Gamma^\alpha{}_{\mu\nu}$ is the one entering into the Galilean connection, and is compatible with $\tau_\mu$ and $h^{\mu\nu}$. The difference of these connections leads to (defining ${^B}\nabla \coloneqq \partial + {^B}\Gamma$):
\begin{align}  
    {^G}\Gamma^\alpha{}_{\mu\nu} - {^B}\Gamma^\alpha{}_{\mu\nu}
        &= - \tau_{(\mu} {^B}\nabla_{\nu)} G^\alpha - \frac{1}{2} h^{\sigma \alpha} \, {^B}\nabla_\sigma  {^G}b_{\mu\nu} \nonumber \\
        &= - \tau_{(\mu} \hat\nabla_{\nu)} G^\alpha - \frac{1}{2} h^{\sigma \alpha} \, \hat\nabla_\sigma  {^G}b_{\mu\nu} + 2 \tau_{(\mu} \kappa_{\nu)\sigma} h^{\sigma \alpha} \nonumber \\
        &= \tau_{\mu} \hat\nabla_{[\nu} B^c \, b_{k] c} + \tau_{\nu} \hat\nabla_{[\mu} B^c \, b_{k] c} + 2 \tau_{(\mu} \kappa_{\nu)\sigma} h^{\sigma \alpha} \nonumber \\
        &= 2 \tau_{(\mu} \left[-\, {^G}\kappa_{\nu) \sigma} + \kappa_{\nu) \sigma}\right] h^{\sigma\alpha},
\end{align}
where we defined ${^G}\kappa_{\mu\nu} \coloneqq {^G}b_{\sigma[\nu} \hat\nabla_{\mu]} G^\sigma$. Finally we get
\begin{align}
    \hat \Gamma^\alpha{}_{\mu\nu}
        &= h^{\mu\sigma}\left(\partial_{(\alpha} \, {^G}b_{\beta)\sigma} - \frac{1}{2} \partial_{\sigma} \, {^G}b_{\alpha\beta} \right) + G^\mu \partial_{(\alpha} \tau_{\beta)} + 2\tau_{(\alpha} {^G}\kappa_{\beta)\nu}  h^{\mu\nu}.
\end{align}
We see that the Galilean connection written as a function of $G^\mu$ has the same form as when written as a function of $B^\mu$, but the Coriolis field is different, i.e. ${^G}\kappa_{\mu\nu} \not= \kappa_{\mu\nu}$.\saut

In conclusion, for a given Galilean connection $\hat \Gamma^\alpha{}_{\mu\nu}$ and a vector $B^\mu$, the Coriolis field~${^B}\kappa_{\mu\nu}$ relative to $B^\mu$ is given by
\begin{align}\label{def_cor}
    {^B}\kappa_{\mu\nu} \coloneqq {^B}b_{\sigma[\nu} \hat\nabla_{\mu]} B^\sigma.
\end{align}
From this formula, when introducing a second affine connection $\geneNR\Gamma^\mu{}_{\alpha\beta}$, the Coriolis field can be re-expressed in the form
\begin{align}
    \kappa_{\mu\nu}
        &= \geneNR\kappa_{\mu\nu} + b_{\sigma[\nu} \left(\tilde \Gamma^\sigma{}_{\mu]\rho} - \hat \Gamma^\sigma{}_{\mu]\rho} \right)B^\rho, \label{kappa_cor}
\end{align}
where we defined the 2-form $\geneNR\kappa_{\mu\nu}$ as
\begin{align}
    \tilde\kappa_{\mu\nu} \coloneqq b_{\sigma[\nu} \tilde\nabla_{\mu]} B^\sigma. \label{cor_fake}
\end{align}
This is a similar definition as~\eqref{def_cor}, but in which the connection involved is not the Galilean connection but the general affine connection. That definition is useful for expressing the distorsion tensor of STENC as a function of the non-metricities of $\tau_\mu$ and $h^{\mu\nu}$ in Sec.~\ref{sec_GR_NR_limit}.

\subsection{Choice of Coriolis field}

\noindent {\it Space-space part}: The difference between the spatial part of two Coriolis fields related to the same Galilean connection is
\begin{align}
    \left({^B}\kappa_{\mu\nu} - {^G}\kappa_{\mu\nu}\right)h^{\mu\alpha} h^{\nu\beta} = D^{[\alpha} v^{\beta]},
\end{align}
where $B^\mu = G^\mu + v^\mu$, with $v^\mu \tau_\mu \coloneqq 0$ and $D^\alpha$ is the spatial connection, i.e. induced by $h^{\mu\nu}$. This formula shows that the freedom on choosing the spatial Coriolis field is related to the gradient of a vector. In other words, we can always find a vector $B^\mu$ such that the 2-form ${^B}\kappa_{\mu\nu}h^{\mu\alpha} h^{\nu\beta}$ has no exact part in his Hodge decomposition. Therefore, the non-exact part of the spatial 2-form Coriolis field is gauge invariant. With this regards, the non-relativistic limit of the Einstein equation constrains this non-exact, gauge-invariant part to be zero.\saut\saut

\noindent {\it Time-space part}: 
We have
\begin{align}
    {^B}\kappa_{\mu\nu} B^\mu h^{\nu\alpha} = \frac{1}{2} {^B}a^\alpha,
\end{align}
where ${^B}a^\alpha \coloneqq B^\mu \hat\nabla_\mu B^\alpha$ is the 4-acceleration of $B^\mu$. So we can always choose a (geodesic) vector $B^\mu$ such that the time-space part of the Coriolis field related to that vector is zero. However, in general, this cannot be done in the same time as removing the exact part of the space-space components of the Coriolis field. In other words, unless a specific physical solution is considered, there does not exist geodesic observers whose Coriolis field is purely non-exact, or, more generally, there does not exist observers whose Coriolis field is zero.


\section{Alternative derivation of the non-relativistic distorsion tensor}\label{app_derivation}

Here, we derive Eq.~\eqref{true_distortion} expressing the non-relativistic distorsion (mostly) in terms of the non-metricities of $\tau_\mu$ and $h^{\mu\nu}$. We recall that $\hat Q_{\mu\nu} \coloneqq \tilde\nabla_\mu \tau_\nu$ and $\hat Q_\mu{}^{\alpha\beta} \coloneqq \tilde\nabla_\mu h^{\alpha\beta}$.
We have the following identities:
\begin{align}
     b_{c(\mu}\hat Q_{\nu)}{}^{c\rho} 
        &\coloneqq b_{c(\mu}\tilde\nabla_{\nu)} h^{c\rho} \nonumber \\
        &= \tilde\nabla_{(\mu}\left(\delta^\rho_{\nu)}-\tau_{\nu)} B^\rho\right) - h^{\rho c} \tilde\nabla_{(\mu} b_{\nu) c} \nonumber \\
        &= -B^\rho \tilde\nabla_{(\mu} \tau_{\nu)} - \tau_{(\mu} \tilde\nabla_{\nu)} B^\rho - h^{\rho c} \tilde\nabla_{(\mu} b_{\nu) c}\, , \\
    B^\rho B^\sigma \tau_{(\mu} \hat Q_{\nu)\sigma}
        &\coloneqq B^\rho B^c \tau_{(\mu} \tilde\nabla_{\nu)} \tau_c \, , \\
   \frac{1}{2} h^{\rho k} b_{c\mu} b_{\nu d} \hat Q_k{}^{cd}
        &\coloneqq \frac{1}{2} h^{\rho k} b_{c\mu} b_{\nu d} \tilde\nabla_k h^{cd} \nonumber\\
        &= \frac{1}{2} h^{\rho k} b_{c\mu} \tilde\nabla_k\left( \delta^c_\nu - \tau_\nu B^c\right) - \frac{1}{2} h^{\rho k} b_{c\mu} h^{cd} \tilde\nabla_k b_{\nu d} \nonumber\\
        &= \frac{1}{2} h^{\rho k}\left[-b_{\mu c} \tau_\nu \tilde\nabla_k B^c - \left(\delta^d_\mu - \tau_\mu B^d\right)\tilde\nabla_k b_{\nu d}\right] \nonumber\\
        &=  h^{\rho k}\left[-b_{c (\mu} \tau_{\nu)} \tilde\nabla_k B^c - \frac{1}{2}\tilde\nabla_k b_{\mu\nu}\right].
\end{align}
The sum of these three terms gives
\begin{align}
    \Sigma^\rho_{\mu\nu} &\coloneqq b_{c(\mu}\hat Q_{\mu)}{}^{c\rho} - B^\rho B^\sigma \tau_{(\mu} \hat Q_{\nu)\sigma} - \frac{1}{2} h^{\rho k} b_{c\mu} b_{\nu d} \hat Q_k{}^{cd} \nonumber\\
        &= - h^{\rho k} \left(\tilde\nabla_{(\mu} b_{\nu) k} - \frac{1}{2}\tilde\nabla_k b_{\mu\nu}\right) - B^\rho \tilde\nabla_{(\mu} \tau_{\nu)} -\tau_{(\mu} \tilde\nabla_{\nu)} B^\rho - B^\rho B^c \tau_{(\mu} \tilde\nabla_{\nu)} \tau_c + h^{\rho k} b_{c (\mu} \tau_{\nu)} \tilde\nabla_k B^c \nonumber \nonumber\\
        &= - h^{\rho k} \left(\tilde\nabla_{(\mu} b_{\nu) k} - \frac{1}{2}\tilde\nabla_k b_{\mu\nu}\right) - B^\rho \tilde\nabla_{(\mu} \tau_{\nu)} -\left[\tau_{\mu} \tilde\nabla_{[\nu} B^c b_{k] c} + \tau_{\nu} \tilde\nabla_{[\mu} B^c b_{k] c}\right] h^{k\rho},
\end{align}
where we recall that $\tilde\kappa_{\mu\nu} \coloneqq b_{\sigma[\nu} \geneNR\nabla_{\mu]} B^\sigma$ and so the final term can be expressed as $2 \tau_{(\mu}\tilde\kappa_{\nu) k} h^{k\rho}$. We see that the first two terms of the above expression enter into the distorsion tensor~\eqref{eq_erfzf}. Consequently, we have
\begin{align}
    \hat{L}^\alpha{ }_{\mu \nu} 
        &= b_{\sigma(\mu}\hat Q_{\nu)}{}^{\sigma\alpha} - B^\alpha B^\sigma \tau_{(\mu} \hat Q_{\nu)\sigma} - \frac{1}{2} h^{\alpha \lambda} b_{\sigma\mu} b_{\nu \rho} \hat Q_{\lambda}{}^{\sigma\rho} + 2\tau_{(\mu}\geneNR\kappa_{\nu)\sigma} h^{\sigma\alpha} -\tau_\mu \tau_\nu h^{\alpha \gamma} \partial_\gamma \phi.
\end{align}
See \cite{Schwartz_2024aee} for further details, discussion, and elaboration on non-relativistic distortion.

\newpage
\nocite{apsrev41Control}
\bibliographystyle{apsrev4-1}
\bibliography{refs}

\end{document}